\documentclass[10pt,showpacs,aps,prd,nofootinbib,floatfix,twocolumn,superscriptaddress,groupedaddress]{revtex4-2}
\usepackage{ragged2e}
\usepackage{amsmath,amssymb}
\usepackage{mathrsfs}
\usepackage{graphicx}
\usepackage{float}
\usepackage{caption}
\usepackage{subcaption}
\usepackage[dvipsnames]{xcolor}
\usepackage{dsfont}
\usepackage{hyperref}
\usepackage[noabbrev]{cleveref}
\hypersetup{colorlinks=true, linkcolor=blue, citecolor=blue}
\usepackage{orcidlink}

\begin{document}

\title{Effects of virtual Majorana neutrinos on charged Lepton Flavor Violation decays from a seesaw variant with radiatively induced light neutrino masses}

\author{Enrique Ram\'irez\orcidlink{0000-0002-8635-4621}$^{(a)}$}
\author{H\'ector Novales-S\'anchez\orcidlink{0000-0003-1503-3086}$^{(a)}$}
\author{Humberto V\'azquez-Castro\orcidlink{0009-0006-7830-9114}$^{(a)}$}
\author{M\'onica Salinas\orcidlink{0000-0001-6413-9552}$^{(b)}$}\email{ salinas\_mon@outlook.com}

\affiliation{
$^{(a)}$Facultad de Ciencias F\'isico Matem\'aticas, Benem\'erita Universidad Aut\'onoma de Puebla, Apartado Postal 1152 Puebla, Puebla, M\'exico\\$^{(b)}$Departamento de F\'isica, Centro de Investigaci\'on y de Estudios Avanzados del IPN, Apartado Postal 14-740, 07000 Ciudad de M\'exico, M\'exico}

\begin{abstract}
Lepton flavor violating decays $\ell_{\alpha} \to \ell_{\beta} \gamma$, being forbidden in the Standard Model framework, provide a sensitive probe for new physics. We study these processes in a seesaw variant in which small neutrino masses are generated radiatively. By analyzing the parameter space constrained by electroweak precision data, we investigate the correlation between these decays and non-unitary effects from TeV-scale heavy neutrinos. According to our results, $\mu \to e \gamma$ is the most promising channel for new physics searches, with the bound $|\eta_{\mu e}| \lesssim 10^{-6}$ obtained for non-unitary effects in this radiative seesaw variant. Our estimations of $\mathcal{BR} \left( \mu \to e \gamma \right)$, which depends on the mass of the heavy neutrinos, shows that both current and future experimental facilities might be sensitive to these effects.
\end{abstract}

\maketitle
\section{Introduction}

All of our knowledge about elementary particles and their interactions has been successfully condensed into a single field theory, known as the Standard Model (SM) of Particle Physics~\cite{Glashow:1961tr,Salam:1968rm,Weinberg:1967tq}. This formulation includes, as a main ingredient, the Brout-Englert-Higgs mechanism, which describes the spontaneous breaking of the electroweak gauge symmetry ${\rm SU}(2)_L \otimes {\rm U}(1)_Y$ into the electromagnetic group ${\rm U}(1)_e$~\cite{Englert:1964et,Higgs:1964pj} by introducing the so-called Higgs field, responsible for generating the masses for all known elementary particles. The Higgs particle, associated with this scalar field, remained elusive for nearly 50 years, until its detection by the ATLAS and CMS Collaborations at the CERN~\cite{ATLAS:2012yve,CMS:2012qbp}, after which the the measurement of the SM particle content was finally complete.
\\

Even though the SM is powerful in its predictions, it is not perfect and, moreover, leaves many unanswered questions, one of which is related to the masses of the neutrinos, as these elementary particles are assumed to be massless in the SM. However, observations from the Super-Kamiokande~\cite{Super-Kamiokande:1998kpq} and the Sudbury Neutrino Observatory~\cite{SNO:2002tuh} experiments demonstrated the existence of neutrino oscillations~\cite{Pontecorvo:1957cp}, which proves that neutrinos must be massive. Because of this, new physics (NP) models have been proposed to understand the origin of these masses. The simplest approach is to assume that neutrinos are Dirac particles, acquiring their masses in the same way as the rest of the fermions in the SM Yukawa sector~\cite{Mohapatra:1998rq}. Nevertheless, to explain the tininess of neutrino masses in this framework, their associated Yukawa couplings would need to be of order $10^{-12} $, which seems ``unnatural''. To address this issue, various NP theories have been proposed, such as the seesaw mechanism~\cite{Mohapatra:1979ia, Mohapatra:1980yp, Pati:1974yy}, models with extra dimensions~\cite{Randall:1999ee, Dienes:1998sb, Arkani-Hamed:1999ylh}, and radiative mass models~\cite{Zee:1980ai,Gelmini:1980re}, among others.
\\

In particular, seesaw models, which are good candidates for explaining the generation of neutrino masses, propose that neutrinos are Majorana particles~\cite{Majorana:1937vz}, with their small masses being related to some high-energy scale $\Lambda$ corresponding to NP beyond the SM. In this context, neutrino masses are $m_{\nu} \sim \frac{v^2}{\Lambda }$, where $v$ is the vacuum expectation value of the SM Higgs potential. Besides these 3 ``light neutrinos'', 3 further heavy neutral leptons, also called ``heavy neutrinos'', emerge with masses $M_{N} \sim \Lambda$. Up to these days, the masses of light neutrinos have not been measured, but, according to current experimental limits~\cite{KATRIN:2021uub,eBOSS:2020yzd,Planck:2018vyg,KATRIN:2024cdt}, they are in the sub-eV range, which would require the NP scale $\Lambda$ to be of the order of grand-unification scales; this fact implies that heavy neutrinos would have masses so large that they are technologically inaccessible, since both their direct production and contributions to low-energy physical processes would be greatly suppressed. This issue of the conventional seesaw mechanism has led to the realization of variants, such as the inverse seesaw~\cite{Mohapatra:1986bd, Gonzalez-Garcia:1988okv, Deppisch:2004fa} and the linear seesaw~\cite{Akhmedov:1995ip, Akhmedov:1995vm}, which are able to overcome this inconvenience. Further, mechanisms to generate neutrino masses radiatively, based on large set of possible diagram topologies, have been realized and implemented in new-physics models~\footnote{A nice and comprehensive review on loop neutrino-mass mechanisms is found in Ref.~\cite{Cai:2017jrq}.}. Diagram topologies yielding neutrino mass have been systematically analyzed and classified at 1 loop~\cite{Bonnet:2012kz} and also at the 2-loop level~\cite{AristizabalSierra:2014wal}. A remarkable instance is, doubtless, the scotogenic model~\cite{Ma:1998dn,Ma:2006km}, an ultraviolet completion of the Weinberg operator~\cite{Weinberg:1979sa} in which Majorana-neutrino masses emerge at the one-loop level and where two dark-matter candidates are introduced. Also, 1-loop Majorana-neutrino masses are generated in another realization of the Weinberg operator, the so-called the Zee model~\cite{Zee:1980ai}. Another model of 1-loop neutrino-mass generation is the minimal radiative inverse seesaw, devised by the authors of Ref.~\cite{Dev:2012sg}. Investigations providing mechanisms and models in which neutrino masses arise radiatively beyond the 1-loop level are also available~\cite{Zee:1985id,Babu:1988ki,Babu:1988ig,Branco:1988ex,Aoki:2008av,Gustafsson:2012vj,Krauss:2002px}. 
\\

In this work, we rather consider the neutrino-mass model presented in reference~\cite{Pilaftsis:1991ug}, in which a condition causing the cancellation of the masses of light neutrinos at tree level, without affecting the masses of heavy neutrinos, is introduced\footnote{The cancellation of tree-level neutrino masses as an outcome of symmetry is discussed in Ref.~\cite{Kersten:2007vk}.}. This new condition yields a break down of the previously mentioned relationship between the masses of heavy and light neutrinos, thus resulting in heavy-neutrino masses untied to some huge energy scale $\Lambda$, then bringing the effects of this NP within the reach of near-future, and even nowadays existing, experimental facilities. On the other hand, the masses of the light neutrinos are rather generated through radiative corrections, providing a natural way to obtain small masses for these particles, as long as heavy-neutrino masses are nearly degenerate.
\\

Within the SM, charged Lepton Flavor Violation (cLFV) processes are forbidden. However, neutrino oscillations imply that such processes are allowed as NP phenomena. Nevertheless, these processes experience a Glashow-Illiopoulos-Maiani (GIM)-like suppression~\cite{Glashow:1970gm} in their loop-level contributions within the SM with massive neutrinos, regardless of whether neutrinos are Dirac or Majorana particles ~\cite{Cheng:1980tp,Petcov:1976ff,Bilenky:1977du,Marciano:1977wx,Lee:1977qz}. Therefore, the consideration of new physics, beyond the SM, seems to be in order. In particular, the cLFV 2-body decays $\ell_{\alpha} \to \ell_{\beta} \gamma$ with $\ell_{\alpha} = \mu, (\tau)$ and $\ell_{\beta} = e,(e,\mu)$, have been studied in various new-physics frameworks; we can mention, for instance,  Refs.~\cite{Kuno:1999jp,Calibbi:2017uvl,Lee:2013htl,Cirigliano:2021img,Hernandez-Tome:2019lkb,Ilakovac:1994kj,Alonso:2012ji,Dudenas:2022von,Xing:2020ivm,Kubo:2006yx,AristizabalSierra:2008cnr,Hisano:1996qq,Casas:2001sr,Davidson:2020hkf,Novales-Sanchez:2016sng,Novales-Sanchez:2017crc,Gonzalez-Quiterio:2024thp}, and references therein. In this work, we consider virtual neutrinos, both heavy and light ones, that induce non-unitary effects in the light neutrino mixing matrix, as discussed in Refs.~\cite{Fernandez-Martinez:2016lgt,Blennow:2023mqx}. The presence of heavy neutrinos and non-unitary effects means that cLFV processes are no longer protected by the GIM mechanism, thereby allowing measurable decay rates for the processes 
$\ell_{\alpha} \to \ell_{\beta} \gamma$. 
\\

Several experimental collaborations have been actively searching for signatures of the cLFV decays $\ell_{\alpha} \to \ell_{\beta} \gamma$. Currently, the most stringent upper limit on the branching ratio of $\mu \to e \gamma$, $\mathcal{BR}(\mu \to e \gamma) \lesssim 10^{-13}$, has been set by the MEG experiment~\cite{MEGII:2025gzr}. In the tau sector, the BaBar~\cite{BaBar:2009hkt} and Belle~\cite{Belle:2021ysv} Collaborations have established less restrictive bounds, with $\mathcal{BR}(\tau \to e \gamma)$ and $\mathcal{BR}(\tau \to \mu \gamma)$ both constrained as $\lesssim 10^{-8}$. Future searches by MEG II~\cite{MEGII:2018kmf} and Belle II~\cite{Belle-II:2022cgf} are expected to improve these limits by approximately 1 order of magnitude for the three processes. In our analysis, we find that among the three cLFV radiative decays, the $\mu \to e \gamma$ process is the most promising. Its branching ratio, considered as a function of the heavy neutrino mass, can reach both current experimental bounds and projected future sensitivity limits. Furthermore, we obtain the constraint $|\eta_{\mu e}| \lesssim 10^{-6}$ for the non-unitary parameter, which is consistent with electroweak precision measurements.
\\

This work is divided into the following sections: in Section \ref{teo} a review of a model of neutrino masses beyond the Standard Model is introduced with all the relevant expressions for the calculation to be carried out later; in Section \ref{deca} details of the relevant decay amplitudes and widths for the cLFV decays at the one-loop level are given; our numerical estimations and analyses of the contributions are developed in Section \ref{resul}; and, finally, we present our conclusions in Section \ref{con}. 

\section{The neutrino model}\label{teo}
Regarding neutrino mass generation, the seesaw mechanism provides a well-established framework, originally developed to explain parity violation as a spontaneously broken symmetry~\cite{Mohapatra:1979ia}. This mechanism predicts the existence of heavy partners $N_j$, known as ``heavy neutral leptons'' or ``heavy neutrinos'', characterized by electric neutrality and large masses, in contrast to the light neutrinos we observe.\\

When $\Lambda$ represents the high-energy scale of spontaneous symmetry breaking where Majorana mass terms emerge, light neutrino masses scale as $m_\nu \sim \frac{v^2}{\Lambda}$, while heavy neutrino masses follow $M_N \sim \Lambda$, with $v = 246\,\mathrm{GeV}$ being the Higgs vacuum expectation value. This establishes the characteristic seesaw relation $m_{N} \sim \frac{v^2}{m_{\nu}}$ between light and heavy neutrino masses. In this context, current experimental bounds on light neutrino masses~\cite{eBOSS:2020yzd,Planck:2018vyg,KATRIN:2024cdt} require an extremely large $\Lambda$ scale, making heavy neutrinos experimentally inaccessible through both direct production and virtual effects. Therefore, seesaw variants have then been developed, each with distinctive assumptions, including the inverse seesaw~\cite{Mohapatra:1986bd,Gonzalez-Garcia:1988okv,Deppisch:2004fa} and linear seesaw~\cite{Akhmedov:1995ip,Akhmedov:1995vm}. \\

For our analysis, we employ the seesaw variant proposed in Ref.~\cite{Pilaftsis:1991ug}, where light neutrino masses vanish at tree level, then being generated through radiative corrections to the neutrino two-point function. Unlike the inverse and linear seesaw mechanism, where the mass relation $m_{N} \sim \frac{v^2}{m_{\nu}}$ gets weakened through additional low-energy scales, this model eliminates the tree-level relation entirely. Instead, it requires a quasi-degenerate heavy neutrino spectrum to generate small radiative masses for light neutrinos. Additional theoretical details supporting our approach can be found in Refs.~\cite{Martinez:2022epq, Novales-Sanchez:2023ztg, Novales-Sanchez:2024pso}.\\

Charged currents, which feature the SM $W$ boson, are taken into consideration when calculating the one-loop contributions to the cLFV processes. The corresponding Lagrangian is
\begin{eqnarray}\label{eq:1}
&&
{\cal L}_{W\nu \ell}=\sum_{\alpha=e,\mu,\tau}\sum_{j=1}^3\frac{g}{\sqrt{2}}
\big(
{\cal B}_{\alpha\nu_j}W^-_\rho\overline{\ell_\alpha}\gamma^\rho P_L\nu_j
\nonumber \\ && \hspace{2cm}
+{\cal B}_{\alpha N_j}W^-_\rho\overline{\ell_\alpha}\gamma^\rho P_LN_j
\big)+{\rm H. c.},
\label{LWnl}
\end{eqnarray}
where $g$ is the ${\rm SU}(2)_L$ coupling constant and $P_L=\frac{1}{2}(\textbf{1}_4-\gamma_5)$ is the left chiral projection matrix. Additionally, $W_\rho$ denotes the SM $W$-boson field, $\ell_ \alpha$ is the SM charged lepton with flavor $\alpha$. The light and heavy mass-eigenspinor Majorana neutrino fields are denoted by $\nu_j$ and $N_j$, respectively. 
The Majorana condition is given by $\psi^c = \psi$, where $\psi^c = C \overline{\psi}^T$ is the charge-conjugated field of $\psi$, and $C$ denotes the charge-conjugation matrix. This condition can only be satisfied if $\psi$ is electrically neutral. In this context, both $\nu_j$ and $N_j$ are Majorana neutrino fields, as they fulfill $\nu_j^c = \nu_j$ and $N_j^c = N_j$. Additionally, $\mathcal{B}_{\alpha \nu_j}$ and $\mathcal{B}_{\alpha N_j}$ are elements of two $3 \times 3$ matrices, denoted as $\mathcal{B}_\nu$ and $\mathcal{B}_N$, respectively. These two matrices together define the following $3 \times 6$ matrix:

\begin{equation}\label{eq:2}
{\cal B}=
\left( 
\begin{array}{cc}
{\cal B}_\nu & {\cal B}_N
\end{array}
\right),
\end{equation} 
which satisfies a pseudo-unitary condition, since ${\cal B}{\cal B}^{\dagger} = \textbf{1}_3$, with $\textbf{1}_3$ representing the $3 \times 3$ identity matrix, but ${\cal B}^{\dagger}{\cal B} = {\cal C}$, where ${\cal C}$ is a Hermitian $6\times6$ matrix that characterizes the deviation from full unitarity. In terms of components, these conditions read

\begin{equation}\label{eq:3}
    \sum_{i=1}^6 {\cal B}_{\alpha n_i} {\cal B}_{\beta n_i}^*= \delta_{\alpha \beta},
\end{equation}

\begin{equation}\label{eq:4}
    \sum_{\alpha = e, \mu, \tau } {\cal B}^*_{\alpha n_j} {\cal B}_{\alpha n_k}= {\cal C}_{jk},
\end{equation}
using $n_i=\nu_1, \nu_2, \nu_3$ for $i=1,2,3$ and $n_i=N_1,N_2,N_3$ for $i=4,5,6$. In addition, the quantities ${\cal C}_{jk}$ are the entries of $\cal C$, as specified in references ~\cite{Pilaftsis:1991ug, Martinez:2022epq, Novales-Sanchez:2024pso,Novales-Sanchez:2023ztg}. 
\\

We expanded the unitary $6\times6$ matrix $\mathcal{U}$, which diagonalizes the mass matrix $\mathcal{M}$ to transition to the mass-eigenstate basis in the neutrino Lagrangian (see~\cite{Pilaftsis:1991ug, Martinez:2022epq, Novales-Sanchez:2023ztg} for details), as a power series in the $3 \times 3$ parameter matrix $\xi = m_\text{D} m_\text{M}^{-1}$~\cite{PhysRevD.47.1080, PhysRevD.86.113001}, where $m_{\rm M}$ and $m_{\rm D}$ are $3\times3$ matrices respectively characterizing Majorana-like and Dirac-like quadratic terms that mix neutrino fields. The expansion assumes $|\xi_{ij}| \ll 1$ to ensure perturbative convergence and avoid unnecessary complexity. The structure of the Majorana and Dirac matrices, $m_\text{M}$ and $m_\text{D}$, is thoroughly discussed in Refs.~\cite{Pilaftsis:1991ug, Martinez:2022epq, Novales-Sanchez:2024pso, Novales-Sanchez:2023ztg}. Let us mention that block matrices comprising $\mathcal{U}$ define $\mathcal{B}_\nu$ and $\mathcal{B}_N$. Consequently, Eq.~\eqref{eq:2} can be reformulated in terms of $\xi$ as follows:

\begin{equation}\label{eq:5}
{\cal B} \simeq
\left( 
\begin{array}{ccc}
 \left( \textbf{1}_3 - \frac{1}{2} \xi \xi^\dagger \right)U^{*}_{\rm PMNS}& & \xi \left( \textbf{1}_3 - \frac{1}{2} \xi^\dagger \xi \right)\hat{V}^{*}
\end{array}
\right),
\end{equation} 
at ${\cal O} (\xi^3)$. Here, $U_{\rm PMNS}$ is the Pontecorvo-Maki-Nakgawa-Sakata (PMNS) matrix~\cite{Maki:1962mu, Pontecorvo:1967fh} and $\hat{V}$ is a $3\times 3$ unitary matrix.
\\

From Equation~\eqref{eq:5}, it is clear that the deviation of the light-neutrino mixing matrix from unitarity is governed by the Hermitian matrix $\xi \xi^{\dagger}$. In the literature, this is often denoted as~\cite{Fernandez-Martinez:2007iaa} 
\begin{equation}
    \eta \equiv \frac{1}{2} \xi \xi^{\dagger},
\end{equation}
where $\eta$ quantifies the non-unitarity effects. Possible phenomenological effects of the non-unitarity neutrino mixing matrix have been explored and discussed in Refs.~\cite{Antusch:2006vwa,Celestino-Ramirez:2024dek,Garnica:2023ccx,Malinsky:2009gw,Dev:2009aw}.

\section{Virtual Majorana neutrino contributions to \texorpdfstring{$\ell_\alpha \to \ell_\beta \gamma$} \text{\hspace{2mm}decays at one loop}} \label{deca}

In this section, we present our analytical calculation of the one-loop contributions to cLFV decays $\ell_\alpha \to \ell_\beta \gamma$, where $\alpha, \beta$ denote charged-lepton flavors with $\alpha = \mu, \tau$ and $\beta = e, \mu$. Of course, we assume that $\beta$ differs from $\alpha$ for flavor violation to take place. Within the theoretical framework of Ref.~\cite{Pilaftsis:1991ug} (summarized in the previous section), these processes receive contributions from both light and heavy Majorana neutrinos propagating in the one-loop Feynman diagrams.
\\

To begin with, neutrinos can be fundamentally described either by Dirac or Majorana fermions, with significant theoretical distinctions between these frameworks. A key difference manifests in their respective Feynman rules, as comprehensively discussed in Refs.~\cite{Denner,Denner2,Gates,Gluza}. While the Majorana description might naively suggest the presence of a larger number of contributing Feynman diagrams, as compared to the Dirac case, careful application of Wick's theorem~\cite{Wick} reveals that the actual set of relevant diagrams contributing to $\ell_\alpha\to\ell_\beta\gamma$ is identical for both fermion types. This equivalence implies that the contributing diagrams cannot distinguish between Majorana and Dirac neutrinos at one loop.
\\

\begin{figure}[H]
\centering
\includegraphics[width=0.3\textwidth]{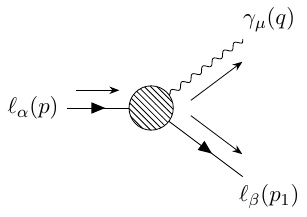}
\caption{\footnotesize \justifying{Conventions for momenta used to carry out the calculation of the $\ell_{\alpha}\to\ell_{\beta}\gamma$ amplitude.}}
\label{fig:generico}
\end{figure}

\begin{figure}[h]
\centering
\begin{subfigure}[h]{0.51\linewidth}
\includegraphics[width=\linewidth]{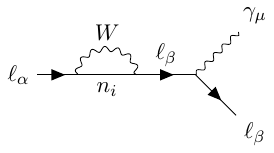}
\caption{}
%\label{fig:westminster_lateral}
\end{subfigure}
\begin{subfigure}[h]{0.51\linewidth}
\includegraphics[width=\linewidth]{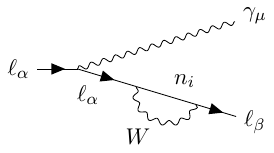}
\caption{}
%\label{fig:westminster_aerea}
\end{subfigure}
\begin{subfigure}[h]{0.51\linewidth}
\includegraphics[width=\linewidth]{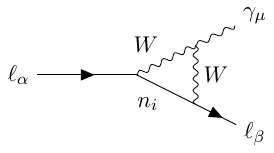}
\caption{}
%\label{fig:westminster_aerea}
\end{subfigure}
\caption{\footnotesize \justifying{Generic Feynman diagrams contributing to the $\ell_\alpha\to\ell_\beta\gamma$ decay amplitudes. Here, $n_{i}$ denotes a neutrino field, either light or heavy, where $n_1=\nu_1$, $n_2=\nu_2$, $n_3=\nu_3$, $n_4=N_1$, $n_5=N_2$  and $n_6=N_3$.}}
\label{fig:diagramas}
\end{figure}

Figure~\ref{fig:generico} illustrates our momenta conventions, where energy-momentum conservation enforces $q = p - p_1$. The complete set of Feynman diagrams contributing to the amplitude in the unitary gauge is presented in Fig.~\ref{fig:diagramas}.

%\begin{figure}[h]
%	\centering
%	\begin{tikzpicture}
%		\begin{feynman}
%			\vertex[blob] (m) at ( 0, 0) {\contour{white}{}};
%			\vertex (a) at (-2,0){$\ell_{\alpha}(p) $};
%			\vertex (b) at (2,1.5){$\gamma_{\mu}(q)$};
%			\vertex (c) at (2,-1.5){$\ell_{\beta}(p_1) $};
			
%			\diagram*{(a) -- [fermion, momentum={}] (m);
%				(m) -- [boson, momentum'={}] (b);
%				(m) -- [fermion, momentum={}] (c);
%			};
%		\end{feynman}
%	\end{tikzpicture}
%	\caption{ Conventions for momenta used to carry out the calculation of the $\ell_{\alpha}\to\ell_{\beta}\gamma$ amplitude.}
%	\label{fig:generico}
%\end{figure}

The amplitude for $\ell_\alpha \to \ell_\beta \gamma$ is given by $i\mathcal{M}=e\overline{u}_{\beta}\Gamma^{\alpha\beta}_{\mu}u_{\alpha}\epsilon^{\mu*}$, where $u_{\alpha}$ and $u_{\beta}$ are momentum-space spinors, the factor $e$ is the electric charge, whereas $\epsilon^{\mu}$ is the polarization vector associated to the electromagnetic field. Under the assumption of Lorentz-invariance, the most general Lorentz-covariant structure for the vertex function characterizing the interaction of two on-shell fermions with a neutral gauge boson, $\Gamma^{\alpha\beta}_{\mu}$, can be written as: 

\begin{equation}\label{eq:6}
\begin{aligned}
    i\Gamma^{\alpha \beta}_{\mu} & = \Big[q^{\nu}\sigma_{\mu \nu} (F^{\alpha \beta}_1+F^{\alpha \beta}_2 \gamma_5)+ \gamma_{\mu}(F^{\alpha \beta}_3+F^{\alpha \beta}_4 \gamma_5) \\
    & + q_{\mu}(F^{\alpha \beta}_5+F^{\alpha \beta}_6 \gamma_5)\Big],
    \end{aligned}
\end{equation}
where the transition form factors $F^{\alpha \beta}_{i}$, $\left( i=1, \ldots ,6\right)$ depend on the squared 4-momentum $q^2$. In this case the vector boson is a photon; therefore, gauge invariance under the gauge group ${\rm U}(1)_e$ implies that the Ward identity~\cite{Ward:1950xp} must be fulfilled. Moreover, for on-shell fermions the momentum-space Dirac equation must be used, resulting in the conditions~\cite{Calibbi:2017uvl}:

\begin{equation}\label{eq:7}
	\begin{aligned}
		&(m_{\alpha}-m_{\beta})F^{\alpha \beta}_3 + q^2F^{\alpha \beta}_5 = 0, \\
		&(m_{\alpha}+m_{\beta})F^{\alpha \beta}_4 + q^2F^{\alpha \beta}_6 = 0.
	\end{aligned}
\end{equation}
For an on-shell photon we have $q^2=0$, so in this case $F^{\alpha \beta}_5$ and $F^{\alpha \beta}_6$, respectively associated to the \textit{scalar} and \textit{pseudo-scalar} form factors, do not play a role in these conditions. On the other hand, $m_{\alpha} \neq m_{\beta}$, so the only solution is $F^{\alpha \beta}_3=F^{\alpha \beta}_4=0$, which means that there are no contributions from the \textit{vector} and \textit{axial} form factors. The decay under consideration only receives contributions from $F^{\alpha \beta}_1$ and $F^{\alpha \beta}_2$, which are associated with the magnetic dipole form factor (MDFF) and the electric dipole form factor (EDFF)~\cite{Hollik:1998vz} respectively. The unpolarized square amplitude for the $\ell_\alpha \to \ell_\beta\gamma$ decays is:
\begin{equation}\label{eq:7a}
        \overline{|\mathcal{M}|}^2 = 8 \pi \alpha_{\textrm{w}} \left(m^2_{\alpha}-m^2_{\beta} \right)^2\left( |F^{\alpha \beta}_1|^2 + |F^{\alpha \beta}_2|^2   \right).
    \end{equation}
With Eq.~\eqref{eq:7a} at hand, the partial
decay width for the cLFV process $l_\alpha\to l_\beta\gamma$ is calculated~\cite{Cheng:1980tp}:
\begin{equation}\label{eq:8}
	\Gamma \left( \ell_{\alpha} \to \ell_{\beta} \gamma \right) = \frac{\alpha_{\rm w} (m^{2}_{\alpha}-m^{2}_{\beta})^3}{2m^3_{\alpha}} \left(|F^{\alpha \beta}_1|^2+|F^{\alpha \beta}_2|^2 \right),
\end{equation}
where $\alpha_{\rm w} = e^2/4 \pi$ is the fine structure constant, and $m_\alpha$, $m_\beta$ are the masses of the charged leptons. 
\\

To properly handle the ultraviolet divergences arising from the Feynman diagrams contributing to the amplitude, we employ dimensional regularization~\cite{Bollini:1972ui, tHooft:1972tcz}. In this scheme, spacetime dimensions are analytically continued to $D = 4 - \epsilon$, with $\epsilon\to0$, thus modifying the loop integrals as
\[
\int \frac{d^{4}k}{(2\pi)^4} \rightarrow \mu^{4-D}_{\rm R} \int \frac{d^{D}k}{(2\pi)^D},
\]
where $\mu_{\rm R}$ is the renormalization scale (with mass dimension 1), introduced to maintain the correct dimensionality of the integrals. The physical limit is recovered by taking $\epsilon \to 0$ after renormalization. For the analytical computation of the Feynman diagrams, we implement the Passarino-Veltman tensor reduction method~\cite{Passarino, DEVARAJ} using the \textsc{Package-X} software~\cite{Patel_2015,Patel:2016fam}.
\\

The MDFF and the EDFF have the following expressions:

\begin{equation}\label{eq:9}
	\begin{aligned}
	& F^{\alpha \beta}_1 =  - \frac{ \alpha_{\rm w} \left( m_{\alpha}+m_{\beta} \right)}{16\pi s^2_{\rm w} m^2_W} \sum^{6}_{i=1} {\cal B}_{\alpha n_i }{\cal B}^{*}_{\beta n_i } H \left( x_i \right), \\
        & F^{\alpha \beta}_2 = - \frac{ \alpha_{\rm w} \left( m_{\alpha}-m_{\beta} \right)}{16\pi s^2_{\rm w} m^2_W} \sum^{6}_{i=1} {\cal B}_{\alpha n_i }{\cal B}^{*}_{\beta n_i } H \left( x_i \right),
	\end{aligned}
\end{equation}
with the loop function $H(x_i)$ given by

\begin{equation}\label{eq:10}
H(x_i) = \frac{5}{6} - \frac{3x_i - 15x_i^2 - 6x_i^3}{12(1-x_i)^3} + \frac{3x_i^3}{2(1-x_i)^4}\ln x_i,
\end{equation}
where $s_{\rm w} \equiv \sin\theta_{\rm w}$ denotes the sine of the weak mixing angle, $m_W$ is the $W$-boson mass, and $x_i \equiv \frac{m_{n_i}^2}{m_W^2}$ denotes the squared ratio between neutrino masses ($m_{n_i} = \{m_{\nu_i}, M_{N_i}\}$) and the $W$-boson mass. The function $H(x_i)$ consists exclusively of 3-point Passarino-Veltman integrals, as the 2-point integrals cancel in the complete amplitude summation. Since the 3-point Passarino-Veltman scalar functions are finite, according to their superficial degree of divergence, our previous statement means that $H(x_i)$ is ultraviolet finite, and thus the MDFF and the EDFF contributions are free of ultraviolet divergences. This function depends on the neutrino mass spectrum (computational details are provided in Appendix~\ref{Ape}). 
\\

From Eq.~\eqref{eq:9}, it is evident that both the MDFF and the EDFF share the same structural form. They are composed of the flavor coefficients from the matrix $\mathcal{B}$, introduced in Section~\ref{teo}, and the function $H(x_i)$. The key distinction between these form factors lies in their mass dependence: the MDFF is proportional to $(m_\alpha + m_\beta)$, whereas the EDFF is proportional to $(m_\alpha - m_\beta)$.
\\

The matrices $\mathcal{B}_{\nu}$ and $\mathcal{B}_{N}$ defined in Eq.~\eqref{eq:2} are sub-block matrices that describe the neutrino flavor mixing of the light- and heavy-neutrino states, respectively. This implies that, using Eq.~\eqref{eq:3}, we can separate their contributions as:

\begin{equation}\label{eq:10a}
    \sum^{6}_{i=1} {\cal B}_{\alpha n_i} {\cal B}^{*}_{\beta n_i} = 
    \sum^{3}_{i=1} {\cal B}_{\alpha \nu_i} {\cal B}^{*}_{\beta \nu_i} + 
    \sum^{3}_{i=1} {\cal B}_{\alpha N_i} {\cal B}^{*}_{\beta N_i} = \delta_{\alpha\beta}.
\end{equation}
Since the lepton flavor indices fulfill $\alpha \neq \beta$, then $\delta_{\alpha\beta}=0$, so the above equation vanishes, thus leading to the following relation:

\begin{equation}\label{eq:10b}
    \sum^{3}_{i=1} {\cal B}_{\alpha \nu_i} {\cal B}^{*}_{\beta \nu_i} = 
    -\sum^{3}_{i=1} {\cal B}_{\alpha N_i} {\cal B}^{*}_{\beta N_i}.
\end{equation}

Applying the previous result to the general structure of the MDFF and EDFF given in Eq.~\eqref{eq:9}, we obtain
  
\begin{equation}\label{eq:11}
    \begin{aligned}
       & \sum^{6}_{i=1} {\cal B}_{\alpha n_i }{\cal B}^{*}_{\beta n_i } H \left( x_i \right)  = \sum^{3}_{i=1} {\cal B}_{\alpha \nu_i }{\cal B}^{*}_{\beta \nu_i}  H_{\nu} \left( x_i \right) \\ 
        & \hspace{1.4cm}+ \sum^{3}_{i=1} {\cal B}_{\alpha N_i }{\cal B}^{*}_{\beta N_i } H_{N} \left( x_i \right), \\
        &\hspace{1.2cm} = \sum^{3}_{i=1} {\cal B}_{\alpha N_i }{\cal B}^{*}_{\beta N_i } \left( H_{N} \left( x_i \right) -H_{\nu} \left( x_i \right)\right).
    \end{aligned}
\end{equation}
This result is crucial as it allows one to express the MDFF and EDFF in Eq.~\eqref{eq:9} as the difference between heavy- and light-neutrino contributions.
\\

Regarding the matrix $\mathcal{B}$, which is written in terms of the block matrices $\mathcal{B}_\nu$ and $\mathcal{B}_N$ shown in Eq.~\eqref{eq:5}, we can express the products $\sum^{3}_{i=1}{\cal B}_{\alpha \nu_i}{\cal B}^*_{\beta \nu_i}$ and $\sum^{3}_{i=1}{\cal B}_{\alpha N_i}{\cal B}^*_{\beta N_i}$ by keeping only first-order contributions with respect to $\xi \xi^\dagger$. This yields:

\begin{equation}\label{eq:12}
    \begin{aligned}
        \sum^{3}_{i=1}{\cal B}_{\alpha \nu_i}{\cal B}^*_{\beta \nu_i} &= \delta_{\alpha\beta} - \sum^3_{i=1} \xi_{\alpha i} \xi^{*}_{\beta i} = \delta_{\alpha \beta} - 2\eta_{\alpha \beta}, \\
        \sum^{3}_{i=1}{\cal B}_{\alpha N_i}{\cal B}^*_{\beta N_i} &= \sum^{3}_{i=1}\xi_{\alpha i} \xi^{*}_{\beta i} = 2\eta_{\alpha \beta},
    \end{aligned}
\end{equation}

It is important to note that while the full matrix product satisfies $\mathcal{BB}^\dagger = \mathbf{1}_3$, the sub-block matrix products $\mathcal{B}_\nu\mathcal{B}_\nu^\dagger$ and $\mathcal{B}_N\mathcal{B}_N^\dagger$ are not restricted to fulfill analogue conditions. In fact, if the sub-blocks ${\cal B}_\nu$ and ${\cal B}_N$ were separately unitary, the contributions to cLFV branching ratios would be significantly suppressed by some sort of double GIM mechanism. For instance, when the $\xi$ matrix, or even the matrix product $\xi \xi^{\dagger}$, is diagonal, the resulting cLFV branching ratios exhibit substantial suppression. We will revisit this point in the next section.
\\

The total decay width for the $\ell_{\alpha} \to \ell_{\beta} \gamma$ processes can be derived by combining \cref{eq:8,eq:9,eq:10,,eq:11,eq:12}, yielding

\begin{equation}\label{eq:13}
    \Gamma\left(\ell_{\alpha} \to \ell_{\beta} \gamma\right) = C_{\alpha\beta} \left| \sum_{i=1}^{3}\xi_{\alpha i}\xi^{*}_{\beta i}G(x_i,y_i) \right|^2,
\end{equation}
where the prefactor $C_{\alpha\beta}$ is given by:

\begin{equation}\label{eq:19}
    C_{\alpha\beta} = \frac{\alpha^3_{\rm w}(m^2_{\alpha}-m^2_{\beta})^3(m^2_{\alpha}+m^2_{\beta})}{256\pi^2 s^4_{\rm w} m^3_{\alpha}m^4_W}.
\end{equation}
Notably, the decay width depends exclusively on the non-unitary contributions encoded in the matrix product $\xi\xi^\dagger$. As emphasized previously, the form factor $G(x_i,y_i)$ captures the interference between heavy- and light-neutrino contributions, which can be expressed as:

\begin{equation}\label{eq:14}
    \begin{aligned}
        G(x_i,y_i) &= H_N(y_i) - H_\nu(x_i), \\
        x_i &= \frac{m^2_{\nu_i}}{m^2_W}, \quad y_i = \frac{M^2_{N_i}}{m^2_W},
    \end{aligned}
\end{equation}

The branching ratio for the flavor-violating radiative decay $\ell_\alpha \to \ell_\beta\gamma$ can be expressed in terms of the dominant leptonic decay mode $\ell_\alpha \to \ell_\beta\overline{\nu_\beta}\nu_\alpha$ as follows:

\begin{equation}\label{eq:14a}
    \mathcal{BR}(\ell_{\alpha} \to \ell_{\beta}\gamma) = 
    \mathcal{BR}(\ell_{\alpha} \to \ell_{\beta}\overline{\nu_{\beta}}\nu_{\alpha}) 
    \frac{\Gamma(\ell_{\alpha} \to \ell_{\beta}\gamma)}{\Gamma(\ell_{\alpha} \to \ell_{\beta}\overline{\nu_{\beta}}\nu_{\alpha})},
\end{equation}
where, $\mathcal{BR}(\ell_\alpha \to \ell_\beta\overline{\nu_\beta}\nu_\alpha)$ represents the experimentally well-established branching ratio of the primary leptonic decay channel~\cite{RevModPhys.73.151}, $\Gamma(\ell_\alpha \to \ell_\beta\gamma)$ denotes the partial width for the flavor-violating radiative process, and $\Gamma(\ell_\alpha \to \ell_\beta\overline{\nu_\beta}\nu_\alpha)$ corresponds to the partial width of the dominant decay mode. Eq.~\eqref{eq:14a} explicitly separates the SM contribution (left factor) from the NP effects contained in the radiative width ratio (right factor).
\\

The decay width $\Gamma\left(\ell_{\alpha} \to \ell_{\beta} \overline{\nu_{\beta}}\nu_{\alpha} \right)$ corresponds to the SM prediction with massless neutrinos and in the limit $m_{\beta} \to 0$:

\begin{equation}\label{eq:15}
    \begin{aligned}
       \Gamma\left(\ell_{\alpha} \to \ell_{\beta} \overline{\nu_{\beta}}\nu_{\alpha} \right) = \frac{G^2_{\rm F} m^5_{\alpha}}{192 \pi^3}, \hspace{3mm} 
    \end{aligned}
\end{equation}
where $G_{\rm F}$ is the Fermi constant, given by:

\begin{equation}
    G_{\rm F} = \frac{\pi \alpha_{\rm w}}{\sqrt{2}s^2_{\rm w} m^2_W},
\end{equation}
The experimental values of the branching ratios $\mathcal{BR}\left(\mu \to e \overline{\nu_{e}}\nu_{\mu} \right)\approx
 1$, $\mathcal{BR}\left(\tau \to e \overline{\nu_{e}}\nu_{\tau} \right)\approx
0.1782$ and $\mathcal{BR}\left(\tau \to \mu \overline{\nu_{\mu}}\nu_{\tau} \right)\approx
0.1739$ are taken from Ref.~\cite{ParticleDataGroup:2024cfk}. 
\\

In order to verify the consistency of our analytical results, we calculate the contribution to the radiative decay $\mu \to e\gamma$ from the SM minimally extended by 3 Dirac light neutrinos~\cite{Cheng:1980tp,Petcov:1976ff,Bilenky:1977du}, which is carried out by using \cref{eq:13,eq:14a,eq:15}. The calculation begins with the cLFV decay width given in \cref{eq:13}. In the limit where only light neutrinos contribute, the loop function $G(x_i,y_i)$ reduces to $H_\nu(x_i)$, which expands as:

\begin{equation}\label{eq:16}
    H_\nu(x_i) \to \frac{5}{6} - \frac{x_i}{4} + \mathcal{O}(x^2_i), \quad \text{where} \quad x_i \equiv \frac{m^2_{\nu_i}}{m^2_W} \ll 1.
\end{equation}

In this regime, the flavor structure is governed by the PMNS matrix through
\[
\sum_i {\cal B}_{\mu \nu_i} {\cal B}^*_{e \nu_i} = \left( U_{\mathrm{PMNS}} U^{\dagger}_{\mathrm{PMNS}} \right)_{\mu e},
\]
 where $B_{\alpha i}$ denotes the neutrino mixing matrix elements. The unitarity of the PMNS matrix~\cite{Maki:1962mu,Pontecorvo:1967fh} enforces a GIM suppression: the leading $\frac{5}{6}$ term cancels exactly, leaving only the neutrino-mass-dependent contribution:

\begin{equation}\label{eq:17}
    \mathcal{BR}(\mu \to e \gamma) = \frac{3\alpha_{\rm w}}{32\pi} \left| \sum_{i=1}^3 U_{\mu i} U^*_{e i} \, x_i \right|^2 \times \mathcal{BR}(\mu \to e\nu\bar{\nu}).
\end{equation}

For the dominant muon decay channel, $\mathcal{BR}(\mu \to e\nu\bar{\nu}) \approx 1$. Factoring out $\Delta m_{31}^2$ and defining $r \equiv \Delta m_{21}^2 / \Delta m_{31}^2 \approx 0.03$ ~\cite{Super-Kamiokande:1998kpq,Esteban:2020cvm}, we obtain:
\begin{equation}
\mathcal{BR}_{\rm SM}(\mu\to e\gamma)\approx  \frac{3\alpha_{\rm w}}{32\pi} \left(\frac{\Delta m_{31}^2}  {m_W^2} \right)^{2} \left| U_{\mu 2} U_{e2}^* r + U_{\mu 3} U_{e3}^* \right|^2.
\label{eq:BR_SM}
\end{equation}

Using $\Delta m^2_{31} \approx 2.5\times10^{-3}$ eV$^2$, for the atmospheric mass-squared splitting, and the $U_{\mu 2} U_{e2}^*$ and  $U_{\mu 3} U_{e3}^*$ allowed values given in Ref. ~\cite{
Esteban:2020cvm}, this evaluates to:

\begin{equation}
\mathcal{BR}_{\rm SM}(\mu\to e\gamma)\approx 10^{-54},
\label{eq:BR_num}
\end{equation}

where the suppression arises from $\Delta m_{31}^2/m_W^2 \sim 10^{-25}$ and loop factors. 
\\

The dramatic suppression ($\mathcal{BR}_{\mathrm{SM}} \approx 10^{-54}$) compared to current experimental limits ($\mathcal{BR}_{\mathrm{exp}} \lesssim 1.5\times10^{-13}$~\cite{MEGII:2025gzr}) conclusively shows that any observable $\mu \to e\gamma$ signal would require physics beyond the minimal extension of the SM neutrino sector.

\section{Numerical Analysis of cLFV Decays}\label{resul}

Within the framework of the model discussed throughout the paper, we present a numerical analysis of cLFV decays. In Section~\ref{deca}, we have discussed how virtual massive neutrinos contribute to the $\ell_{\alpha} \to \ell_{\beta} \gamma$ decay at the one-loop level.
\\

While cLFV processes have never been measured, extensive searches by experimental collaborations have established stringent upper limits on their branching ratios. Table~\ref{tab:limi} summarizes: current experimental bounds for various cLFV channels and projected sensitivities of future experiments.
\\

\begin{table}[h]
\begin{center}
\begin{tabular}{ c  c  c }
\hline
Decay & Current Bound & Future Sensitivity \\
\hline
\hline
$\mathcal{BR(}\mu \to e \gamma)$ & $1.5 \times10^{-13}$~\cite{MEGII:2025gzr} & $6 \times10^{-14}$~\cite{MEGII:2018kmf} \\
$\mathcal{BR}(\tau \to e \gamma)$ & $3.3 \times10^{-8}$~\cite{BaBar:2009hkt} & $3 \times10^{-9}$~\cite{Belle-II:2022cgf} \\
$\mathcal{BR}(\tau \to \mu \gamma)$ & $4.2 \times10^{-8}$~\cite{Belle:2021ysv} & $ 10^{-9}$~\cite{Belle-II:2022cgf} \\
\hline
\hline
\end{tabular}
\caption{Upper bounds and future sensitivity for $\ell_{\alpha} \to \ell_{\beta} \gamma $ decays at $90 \%$ C.L.}
\label{tab:limi}
\end{center}
\end{table}
Although neutrinos are massless in the original SM formulation, observations of neutrino oscillations by Super-Kamiokande~\cite{Super-Kamiokande:1998kpq} and the Sudbury Neutrino Observatory~\cite{SNO:2002tuh} demonstrated that neutrinos must be massive~\cite{Pontecorvo:1957cp}, providing compelling evidence for physics beyond the SM. While cosmological observations constrain the sum of neutrino masses to $\sum_j m_{\nu_j} \lesssim 0.12$ eV at 95\% C.L. ~\cite{eBOSS:2020yzd,Planck:2018vyg}, the absolute mass scale remains challenging to determine; for Majorana neutrinos, the effective mass $m_{\beta\beta} = \left| \sum_i (U_{\text{PMNS}})_{ei}^2 m_{\nu_i} \right|$ could be probed through the elusive neutrinoless double beta decay~\cite{Schechter:1981bd}, with current null results from CUORE~\cite{CUORE:2019yfd}, GERDA~\cite{GERDA:2020xhi}, and KamLAND-Zen~\cite{KamLAND-Zen:2016pfg} setting upper limits in the $10^{-2}$--$10^{-1}$ eV range. Complementary direct measurements by KATRIN have recently established the upper limit $m_\nu \lesssim 0.45$ eV (90\% C.L.) for the electron antineutrino mass~\cite{KATRIN:2024cdt}, which we adopt as a conservative model-independent bound in our analysis.
\\

In our analysis, based on the model~\cite{Pilaftsis:1991ug}, the heavy-neutrino masses must exhibit near-degeneracy to maintain the smallness of light-neutrino masses. We therefore implement the approximation $M_{N_1} \approx M_{N_2} \approx M_{N_3} \equiv M_N$, with $M_N$ spanning from 10 GeV to 1.5 TeV to cover phenomenologically relevant scales. For the light-neutrino sector, we conservatively adopt the KATRIN upper limit $m_\nu = 0.45$ eV as a benchmark value throughout our calculations.
\\

The remaining free parameters in our framework are the elements of the matrix $\xi\xi^\dagger$, which currently lack direct experimental constraints. However, building on the mass hierarchy established in the previous section ($M_N \gg m_\nu$), we can analyze the form factor $G(x_i,y_i)$ in the asymptotic limits: 
\begin{equation}
\begin{aligned}
H_N(\infty) \equiv &\lim_{M_N\to\infty} H(y_i) = \frac{1}{3}, 
\\
H_\nu(0) \equiv &\lim_{m_\nu\to 0} H(x_i) = \frac{5}{6}.
\end{aligned}
\end{equation}

Using these limits in Eq.~\eqref{eq:13}, we obtain the decay width:
\begin{equation}\label{eq:18}
\begin{aligned}
\Gamma(\ell_\alpha \to \ell_\beta \gamma) &= C_{\alpha\beta} \left| \sum_{i=1}^3 \xi_{\alpha i} \xi_{\beta i}^* \left( H_N(\infty) - H_\nu(0) \right) \right|^2 \\
&= C_{\alpha\beta} \left| -\frac{1}{2} \sum_{i=1}^3 \xi_{\alpha i} \xi_{\beta i}^* \right|^2 \\
&= \frac{C_{\alpha\beta}}{4} \left| \sum_{i=1}^3 \xi_{\alpha i} \xi_{\beta i}^* \right|^2,
\end{aligned}
\end{equation}
where the numerical values of $H_N(\infty)$ and $H_\nu(0)$ emerge from the explicit form of $G(x_i,y_i)$ in the respective limits.
\\

The branching ratio given in Eq.~\eqref{eq:14a}, the amplitude in Eq. ~\eqref{eq:18} and the experimental bounds for the cLFV decays given in Table~\ref{tab:limi} can be used to determine an upper bound on the contribution of the off-diagonal elements of the matrix $\xi\xi^{\dagger}$, in the following way:
\begin{equation}
    \Bigg|\sum_{i=1}^{3}\xi_{\alpha i}\xi_{\beta i}^{*}\Bigg| \approx \sqrt{\frac{8 \pi}{3 \alpha_{\rm w}}\frac{\mathcal{BR}(\ell_\alpha \rightarrow\ell_\beta \gamma)}{\mathcal{BR}(\ell_\alpha \rightarrow\ell_\beta \bar{\nu}_\beta \nu_\alpha)}},
\end{equation}
from which we obtain the bounds
\begin{equation}\label{eq:20}
    \begin{aligned}
       & \big| \eta_{\mu e}\big|= \frac{1}{2}\Bigg| \sum_{i=1}^{3}\xi_{\mu i}\xi^{*}_{e i} \Bigg| \lesssim 6.6 \times10^{-6}, \\ 
       & \big| \eta_{\tau e}\big|= \frac{1}{2}\Bigg| \sum_{i=1}^{3}\xi_{\tau i}\xi^{*}_{e i} \Bigg| \lesssim 7.5\times 10^{-3} ,\\ 
       & \big| \eta_{\tau \mu}\big| = \frac{1}{2}\Bigg| \sum_{i=1}^{3}\xi_{\tau i}\xi^{*}_{\mu i} \Bigg| \lesssim 8.5\times 10^{-3} .\\ 
    \end{aligned}
\end{equation}
It is important to note that the upper limits presented in Eq.~\eqref{eq:20} are derived under the non-general assumption of very large heavy neutrino masses ($M_{N_i} \gg m_W$). In this regime, the branching ratios depend solely on the elements $|\eta_{\alpha \beta}| = \frac{1}{2}|\sum_{i=1}^{3}\xi_{\alpha i}\xi_{\beta i}^{*}|$, allowing each decay channel to be constrained independently. However, this simplification no longer holds when considering heavy neutrinos with masses in the range of 10--1500 GeV, as in our analysis.
\\

The mass range $M_N \in [10, 1500]$ GeV was chosen to focus on the collider-accessible regime, where heavy neutrinos are produced via electroweak processes and are directly constrained by ATLAS and CMS. While the sub-10 GeV region has been extensively studied in the context of rare meson and $\tau$ decays~\cite{Bondarenko_2018, PhysRevD.110.035029, article}, our results remain quantitatively unchanged in that region.\\

In the following sections, we present various parameterizations of the mixing matrix $\xi$ and perform a detailed numerical analysis of the cLFV branching ratios across this mass range.

\subsection{Quasi-diagonal texture}

In previous studies by some of the authors of the present investigation~\cite{Martinez:2022epq,Novales-Sanchez:2023ztg,Novales-Sanchez:2024pso}, the matrix $\xi$ was parameterized by an approximately diagonal structure, that is, with $\xi_{ii} \approx 1$ and small off-diagonal entries $|\xi_{ij}| \ll 1$ ($i \neq j$), scaled by a real parameter $\rho$ and a phase factor $e^{i\phi}$. Although $\xi$ approaches the identity matrix form, the non-vanishing off-diagonal elements - while strongly suppressed - play a crucial role in mediating flavor-violating processes. This parameterization choice originally reflected the weak dependence of earlier results on specific off-diagonal patterns. Our current analysis reveals that this quasi-diagonal structure, particularly when implemented with degenerate heavy neutrino masses as demanded by the model~\cite{Pilaftsis:1991ug}, induces significant suppression of cLFV branching ratios. The matrix structure is explicitly given by:

\begin{equation}\label{eq:21}
    \xi = \rho e^{i\phi}
    \begin{pmatrix}
        1 & \epsilon_{12} & \epsilon_{13} \\
        \epsilon_{21} & 1 & \epsilon_{23} \\
        \epsilon_{31} & \epsilon_{32} & 1
    \end{pmatrix}, 
    \quad \text{where} \quad |\epsilon_{ij}| \ll 1 \quad (i \neq j).
\end{equation}

Using the above equation, the parameterization of the mixing matrix $\mathcal{B}$ in Eq.~\eqref{eq:5} goes as follows:

\begin{equation}\label{eq:22}
{\cal B} \simeq
\left( 
\begin{array}{ccc}
 \left( 1 - \frac{\rho^2}{2}  \right)U^{*}_{\rm PMNS}& & \rho e^{i\phi}\left( 1 - \frac{\rho^2}{2} \right)\hat{V}^{*}
\end{array}
\right).
\end{equation} 
Note that Eq.~\eqref{eq:22} depends solely on the unitary matrices $U_{\text{PMNS}}$ and $\hat{V}$. Consequently, when computing the products $\mathcal{B}_{\nu}\mathcal{B}^{\dagger}_{\nu}$ and $\mathcal{B}_{N}\mathcal{B}^{\dagger}_{N}$, we obtain

\begin{equation}\label{eq:23}
    \begin{aligned}
        \sum^{3}_{i=1} {\cal B}_{\alpha \nu_i} {\cal B}^{*}_{\beta \nu_i} &= \left(1 - \frac{\rho^2}{2}\right)^2 \delta_{\alpha\beta}, \\
        \sum^{3}_{i=1} {\cal B}_{\alpha N_i} {\cal B}^{*}_{\beta N_i} &= \rho^2 \left(1 - \frac{\rho^2}{2}\right)^2 \delta_{\alpha\beta},
    \end{aligned}
\end{equation}
where $\delta_{\alpha\beta}$ is the Kronecker delta. Both expressions vanish for $\alpha \neq \beta$, as expected from the unitarity of the mixing matrices. The contributions from the light and heavy mixing-matrix elements cancel independently, leading to the following relations:

\begin{equation}\label{eq:24}
    \begin{aligned}
        \sum^{2}_{i=1} {\cal B}_{\alpha \nu_i} {\cal B}^{*}_{\beta \nu_i} &= - {\cal B}_{\alpha \nu_3} {\cal B}^{*}_{\beta \nu_3}, \\
        \sum^{2}_{i=1} {\cal B}_{\alpha N_i} {\cal B}^{*}_{\beta N_i} &= - {\cal B}_{\alpha N_3} {\cal B}^{*}_{\beta N_3},
    \end{aligned}
\end{equation}
where the cancellation reflects the unitarity constraints. By exploiting these relations, we can express the contributions of the light and heavy neutrinos to the MDFF and EDFF in Eq.~\eqref{eq:9} by decomposing the sums in Eq.~\eqref{eq:24} as follows:

\begin{equation}\label{eq:25}
    \begin{aligned}
        \sum^{3}_{i=1} {\cal B}_{\alpha \nu_i} {\cal B}^{*}_{\beta \nu_i} H_{\nu}(x_i) &= \sum^{2}_{i=1} {\cal B}_{\alpha \nu_i} {\cal B}^{*}_{\beta \nu_i} \left( H_{\nu}(x_i) - H_{\nu}(x_3) \right), \\
        \sum^{3}_{i=1} {\cal B}_{\alpha N_i} {\cal B}^{*}_{\beta N_i} H_{N}(y_i) &= \sum^{2}_{i=1} {\cal B}_{\alpha N_i} {\cal B}^{*}_{\beta N_i} \left( H_{N}(y_i) - H_{N}(y_3) \right).
    \end{aligned}
\end{equation}
This result, derived from a quasi-diagonal parameterization of the $\xi$ matrix, demonstrates that the electromagnetic form factor contributions cancel separately for light and heavy neutrinos. For light neutrinos, this cancellation occurs because their masses are small and nearly degenerate, while for heavy neutrinos, the cancellation results from their quasi-degenerate mass spectrum. Consequently, the mass-dependent differences in $H_{\nu}(x_i)$ and $H_{N}(y_i)$ become negligible, leading to strongly suppressed branching ratios. In contrast to previous studies~\cite{Martinez:2022epq,Novales-Sanchez:2023ztg,Novales-Sanchez:2024pso}, which employed diagonal parameterizations, for the present work we will consider non-diagonal parameterizations in order to take into account different scenarios in our model.

\subsection{Cross \texorpdfstring{$\xi$}\text{\hspace{2mm}texture}}
The cross texture is characterized by two real parameters $\rho_1$, $\rho_2$ and a complex phase $\phi$, defined through the matrix structure:

\begin{equation}\label{eq:26}
    \xi =  e^{i\phi}
    \begin{pmatrix} 
        \rho_1 & \rho_2 & \rho_1 \\
        \rho_2 & \rho_2 & \rho_2 \\
        \rho_1 & \rho_2 & \rho_1
    \end{pmatrix},
\end{equation}
Now, we can express the mixing matrix $\eta$ as:

\begin{equation}\label{eq:27}
    \begin{aligned}
      2\eta_{\alpha \beta} = \left(\xi \xi^{\dagger} \right)_{\alpha \beta}= \begin{pmatrix}
          2 \rho^2_1 + \rho^2_2 & 2 \rho_1\rho_2 + \rho^2_2 & 2 \rho^2_1 + \rho^2_2 \\
          2 \rho_1\rho_2 + \rho^2_2 & 3\rho^2_2 & 2 \rho_1\rho_2 + \rho^2_2 \\
          2 \rho^2_1 + \rho^2_2 & 2 \rho_1 \rho_2+ \rho^2_2 & 2 \rho^2_1 + \rho^2_2
      \end{pmatrix}_{\alpha \beta}.  
    \end{aligned}
\end{equation}
This parameterization reveals that the elements of the mixing matrix obey the following relations:

\begin{equation}\label{eq:28}
    \begin{aligned}
       &\eta_{ee}=\eta_{e \tau}=\eta_{ \tau e} = \eta_{\tau \tau}, \\
       & \eta_{e \mu} = \eta_{\mu e}= \eta_{\mu \tau}=\eta_{ \tau \mu}.
    \end{aligned}
\end{equation}
These relations constrain the possible values of the mixing matrix components within this framework. Current experimental searches for non-unitary effects in the light-neutrino sector have established upper bounds on the parameters $|\eta_{\alpha \beta}|$ at the level of $10^{-2}$, derived from combined analyses of short-baseline and long-baseline neutrino oscillation experiments~\cite{Forero:2021azc}. Future high-energy experiments are expected to improve the sensitivity to non-unitary effects by approximately one order of magnitude in the light neutrino sector~\cite{Miranda:2020syh,Soumya:2021dmy,Agarwalla:2021owd,Celestino-Ramirez:2023zox}. For our analysis, we adopt the conservative upper bounds on $|\eta_{\alpha \beta}|$ parameters obtained from a global fit of current flavor and electroweak precision observables at $95 \%$ confidence level~\cite{Blennow:2023mqx}:

\begin{equation}\label{eq:29}
    \begin{aligned}
        &|\eta_{e \mu}| \lesssim 1.2 \times 10^{-5}, \\
        &|\eta_{e \tau}| \lesssim 9 \times 10^{-4}, \\
        &|\eta_{\mu \tau}| \lesssim 5.7 \times 10^{-5}.
    \end{aligned}
\end{equation}

In what follows, we refer to the bounds in Eq.~\eqref{eq:29} as those \emph{derived from electroweak data} (E.W.\ data). 
%In the context of cLFV processes such as $\ell_\alpha \to \ell_\beta \gamma$ ($\alpha \neq \beta$), only the off-diagonal entries of $\eta$ contribute directly to the decay amplitudes. 
These were obtained in Ref.~\cite{Blennow:2023mqx}, where the so-called \emph{3N-SS} scenario, i.e.\ the Standard Model extended by three extra sterile neutrinos with a seesaw-like mass generation mechanism, is considered and then a global fit including $m_W$, $\sin^2\theta_{\rm eff}$, $Z$-pole observables, leptonic $W$ decays, cLFV processes, and CKM unitarity tests is performed. 
\\

For the numerical analysis, we consider the parameter ranges $\{\rho_1, \rho_2\} \in [-8, 8] \times 10^{-3}$. The light-neutrino mass is taken to be $m_\nu = 0.45$ eV~\cite{KATRIN:2024cdt}. The relevant physical parameters from the Particle Data Group~\cite{ParticleDataGroup:2024cfk} are summarized in Table~\ref{tab:parameters}:

\begin{table}[H]
\centering
\begin{tabular}{lc}
\hline
Parameter & Value \\
\hline
\hline
$W$ boson mass ($m_W$) & 80.369 GeV \\
Electron mass ($m_e$) & 0.510 MeV \\
Muon mass ($m_\mu$) & 105.65 MeV \\
Tau mass ($m_\tau$) & 1776.93 MeV \\
Fine-structure constant ($\alpha$) & $1/137.035$ \\
\hline
\hline
\end{tabular}
\caption{Physical parameters used in the analysis.}
\label{tab:parameters}
\end{table}
For the tau lepton decays, $\tau \to \ell_{\beta}\gamma$ ($\ell_{\beta} = e, \mu$), the experimental upper bounds on the branching ratios are less stringent compared to the muon decay case (see Table~\ref{tab:limi}). Consequently, in our numerical analysis, we primarily utilize the more restrictive constraints from the $\mu \to e \gamma$ process. This approach is necessary because any viable point in the $\{\rho_1,\rho_2\}$ parameter space must simultaneously satisfy all three experimental bounds.
\begin{figure}[H]
\centering
\includegraphics[width=0.53\textwidth]{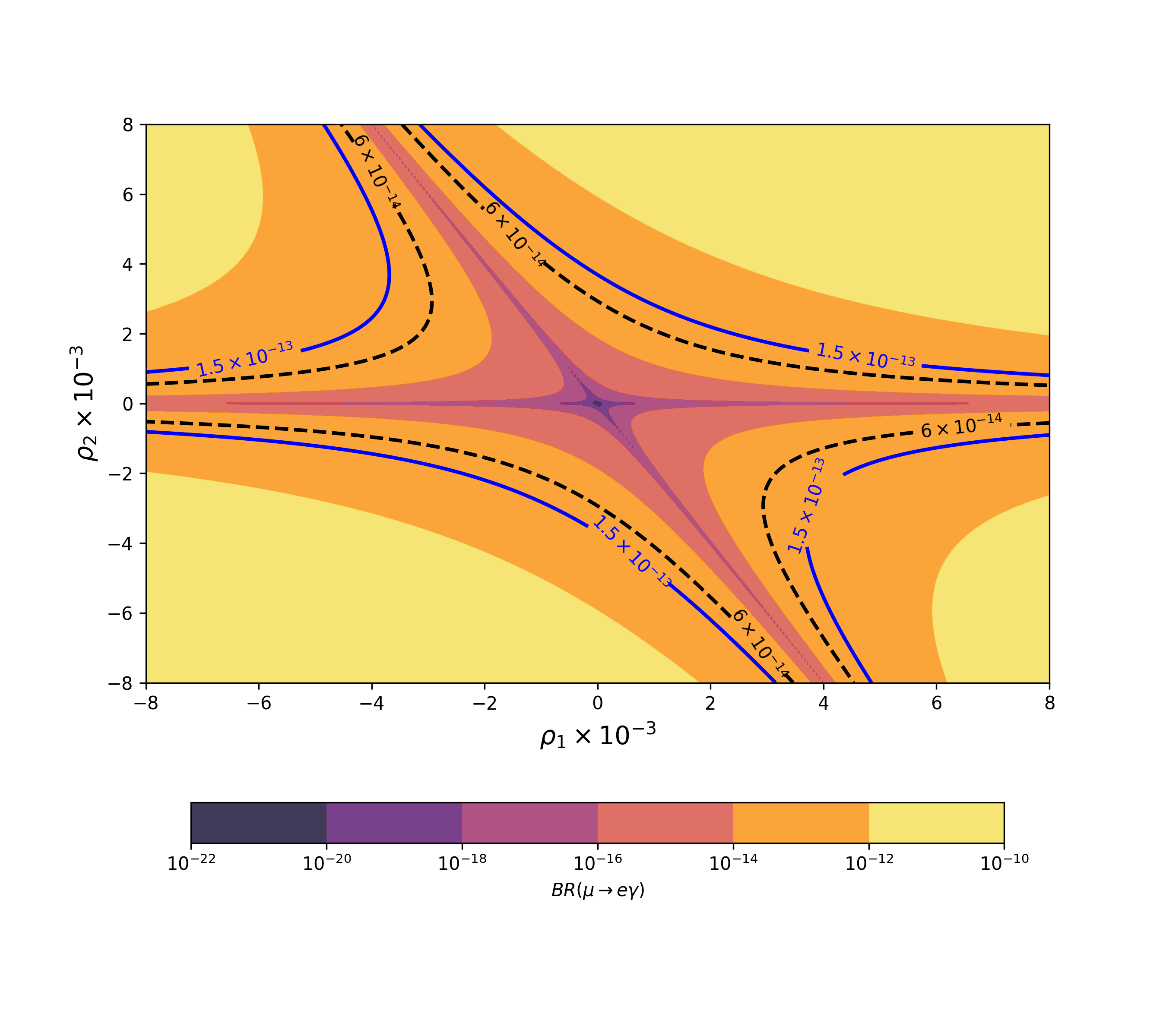}
\caption{\footnotesize \justifying{Branching ratio $\mathcal{BR}(\mu \to e \gamma)$ in the $\rho_1$--$\rho_2$ parameter space. The current experimental upper limit $1.5 \times 10^{-13}$ is represented by the blue solid curve; projected sensitivity, $6 \times 10^{-14}$, is represented by the black dashed curve (see Table~\ref{tab:limi}).}}
\label{fig:espacio}
\end{figure}
Figure~\ref{fig:espacio} shows the allowed parameter space in the $(\rho_1, \rho_2)$ plane, considering the branching ratio constraints from $\mu \to e \gamma$ decay with a heavy virtual neutrino mass of $M_N = 1500$ GeV. The plot reveals two distinct regions: 
\begin{itemize}
    \item The solid blue curve corresponds to the current experimental bound $\mathcal{BR}(\mu \to e \gamma) \lesssim 1.5 \times 10^{-13}$.
    \item The dotted black curve represents the projected future sensitivity of $\mathcal{BR}(\mu \to e \gamma) \lesssim 6 \times 10^{-14}$ (see Table~\ref{tab:limi} for numerical values).
\end{itemize}

Note that ${\cal BR}(\mu\to e\gamma)$ exhibits asymptotic behavior as $\rho_2 \to 0$, implying that $\rho_2$ cannot vanish, while $\rho_1$ remains unconstrained within the considered range $[-8,8] \times 10^{-3}$. The plot reveals a hyperbolic structure comprising two distinct pairs of hyperbolas, both sharing the common asymptote at $\rho_2 = 0$. This characteristic shape indicates an inverse relationship between $\rho_1$ and $\rho_2$ in determining the branching ratio. The observed dependence arises because the branching ratio is proportional to the following combination of parameters:
\begin{equation}
    \mathcal{BR}(\mu \to e \gamma) \propto \rho_2 (2\rho_1 + \rho_2) \times f(m_W, M_N, m_\mu, m_e),
\end{equation}
where $f(m_W, M_N, m_\mu, m_e)$ represents the remaining mass-dependent terms of the function.\\

\begin{figure}[H]
\centering
\includegraphics[scale=0.54]{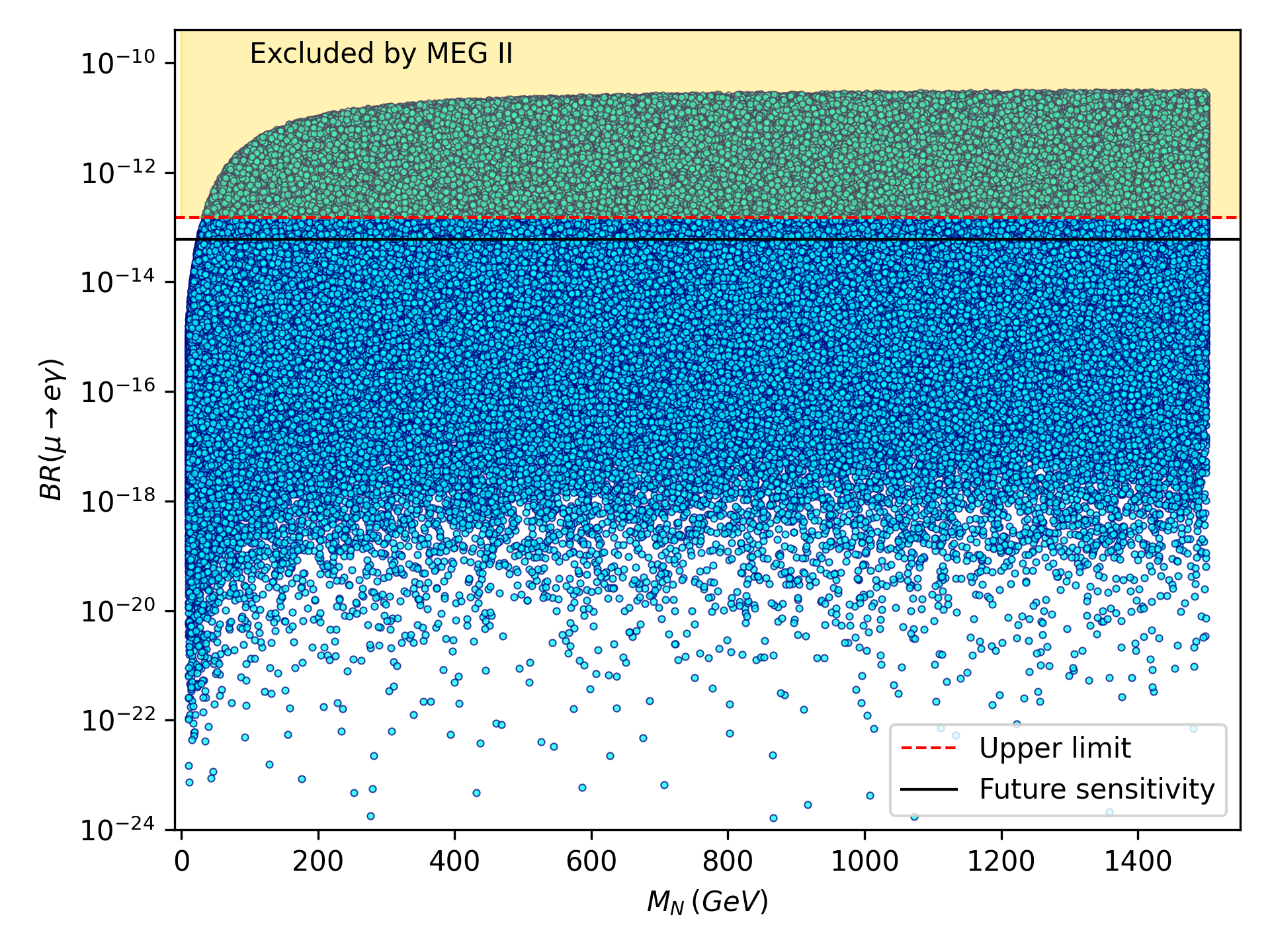}
\caption{ \footnotesize \justifying{ Branching ratio $\mathcal{BR}(\mu \to e \gamma)$ depending on the mass of the heavy neutrino $M_N$ parameter with randomly sampled points. Current experimental upper limit ($1.5 \times 10^{-13}$) is shown as red dashed line; projected sensitivity ($6 \times 10^{-14}$) is given by the black solid line.}}
\label{fig:mue}
\end{figure}

\begin{figure}[h]
	\centering
	\begin{subfigure}{\linewidth}
		\includegraphics[width=\linewidth]{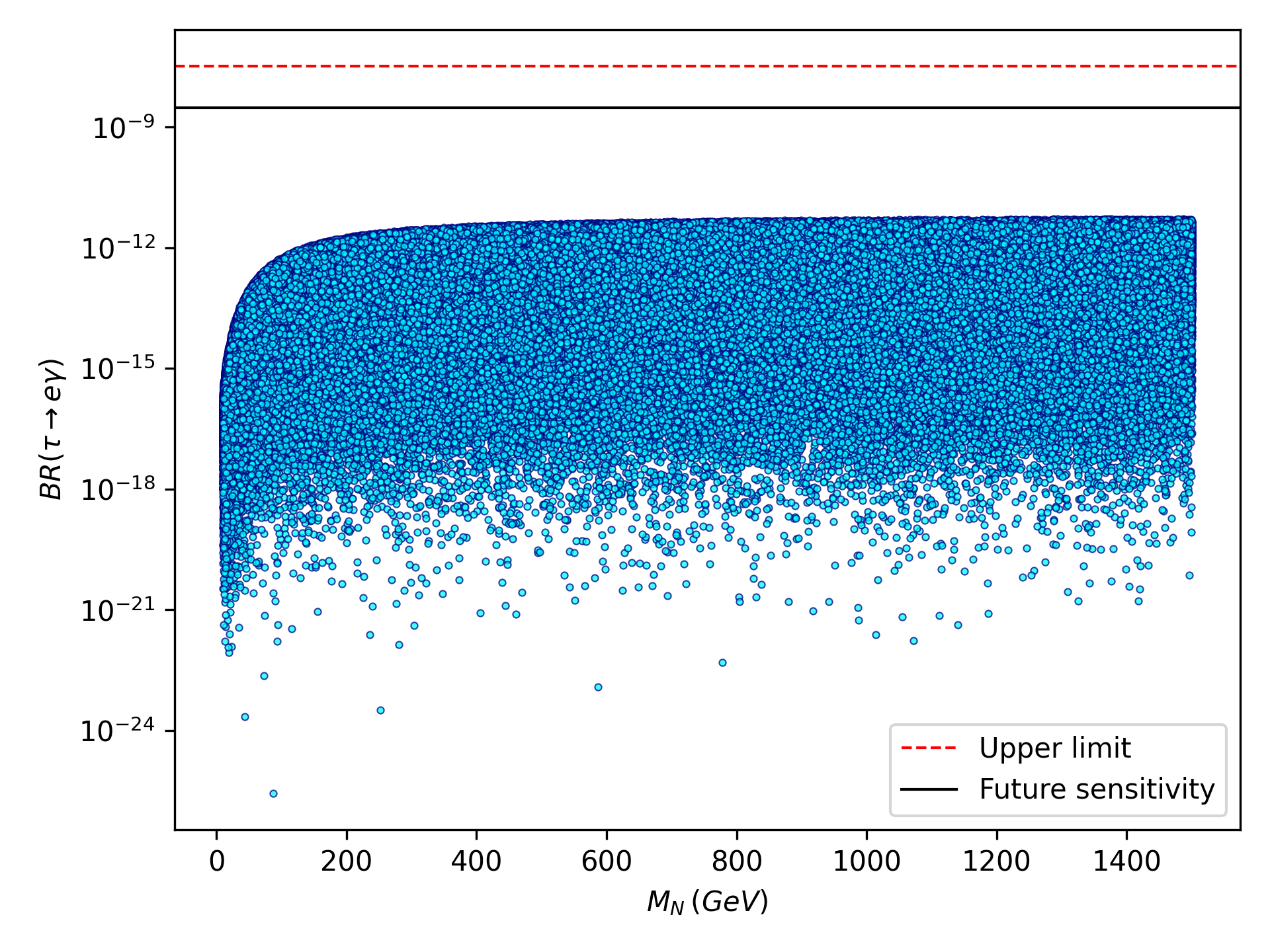}
		\caption{Scatter plot of $BR \left( \tau \to e \gamma \right)$.}
		\label{fig:subfigA}
	\end{subfigure}
    \vfill
	\begin{subfigure}{\linewidth}
		\includegraphics[width=\linewidth]{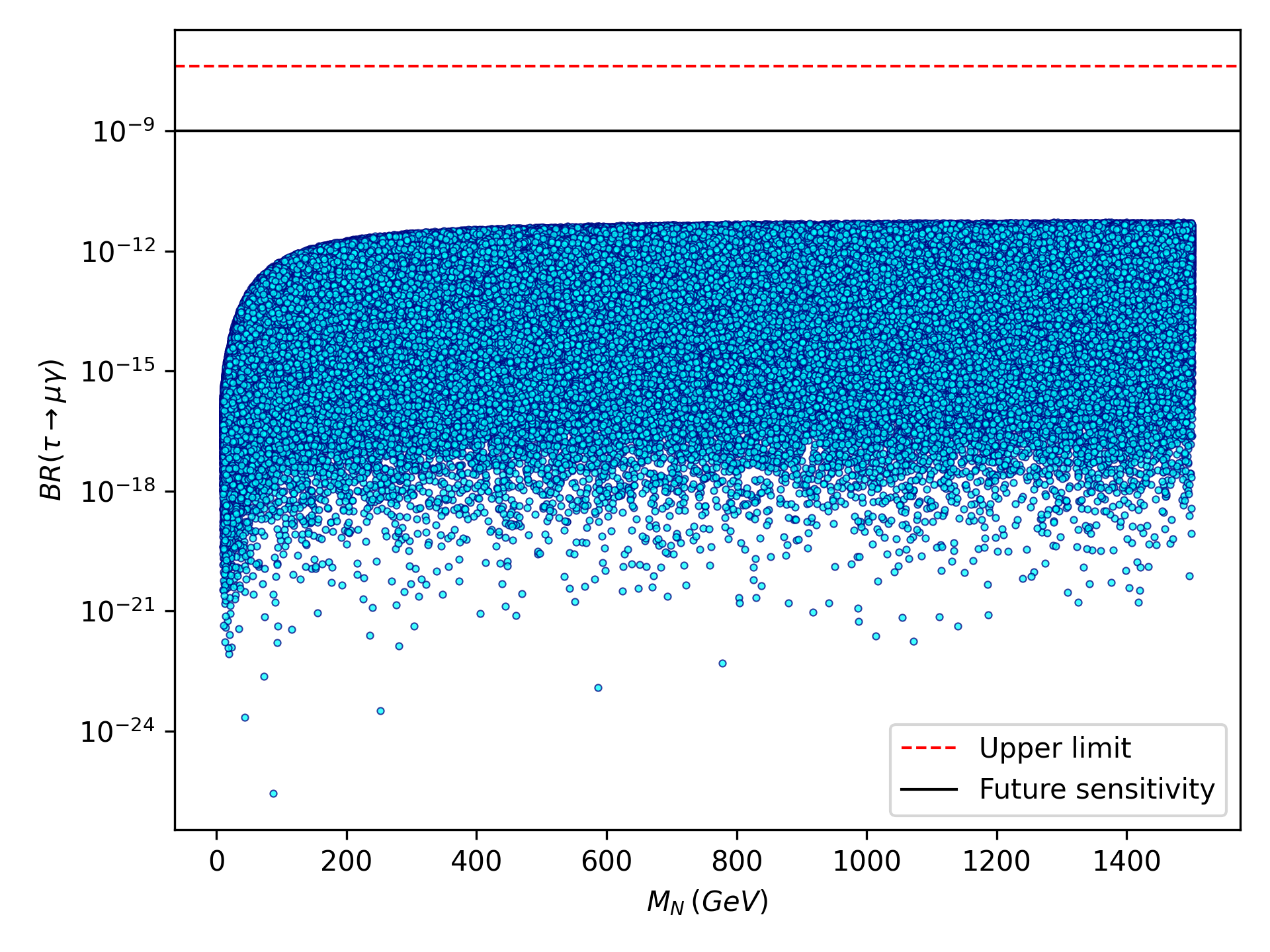}
		\caption{Scatter plot of $BR \left( \tau \to \mu \gamma \right)$.}
		\label{fig:subfigB}
	\end{subfigure}
	\caption{\footnotesize \justifying{Branching ratio $\mathcal{BR}(\tau \to e \gamma)$ and $\mathcal{BR}(\tau \to \mu \gamma)$ depending on the mass of the heavy neutrino $M_N$ parameter with randomly sampled points. The current experimental upper limits are shown as red dashed lines and the projected sensitivities as a black solid lines.}}
	\label{fig:taus}
\end{figure}

Figures~\ref{fig:mue} and \ref{fig:taus} display the branching ratios of the $\ell_{\alpha} \to \ell_{\beta} \gamma$ decays as functions of the heavy-neutrino mass $M_N$, along with current and future experimental limits (see Table~\ref{tab:limi}). During the numerical evaluation, a random scan was taken for the masses of heavy neutrinos within the range $[10,1500]$ GeV, while the parameters $\rho_{1,2}$ took random values within the range $[-8,8]\times 10^{-3}$. For the $\mu \to e \gamma$ decay channel (Fig.~\ref{fig:mue}), the yellow band indicates the parameter space excluded by the MEG II experiment~\cite{MEGII:2025gzr}, while the region below the red dashed line remains consistent with current experimental bounds. The black solid line represents the projected sensitivity of MEG II~\cite{MEGII:2018kmf}, with the area between the red and black lines corresponding to the parameter space accessible to future experimental searches. 

 The $\tau \to \ell_\beta \gamma$ decays ($\beta = e, \mu$), shown in Fig.~\ref{fig:taus}, are suppressed by at least four orders of magnitude relative to both current and future experimental sensitivities, rendering them unobservable at Belle~II~\cite{Belle-II:2022cgf}. This suppression arises from the stringent constraints imposed by the $\mu \to e \gamma$ channel, as discussed above. All three radiative decays exhibit a similar dependence on $M_N$ across the mass range $[10, 1500]~\mathrm{GeV}$. This common behavior originates from the universal structure of Eq.~\eqref{eq:14a}, while variations, ocurring in each specific decay process, emerge from the elements of the mixing matrix $\xi \xi^\dagger$ and the branching ratios of the primary leptonic decays $\mathcal{BR}(\ell_\alpha \to \ell_\beta \overline{\nu}_\beta \nu_\alpha)$.
 
\begin{widetext}

\begin{figure}[h]
 \centering
\includegraphics[scale=0.6]{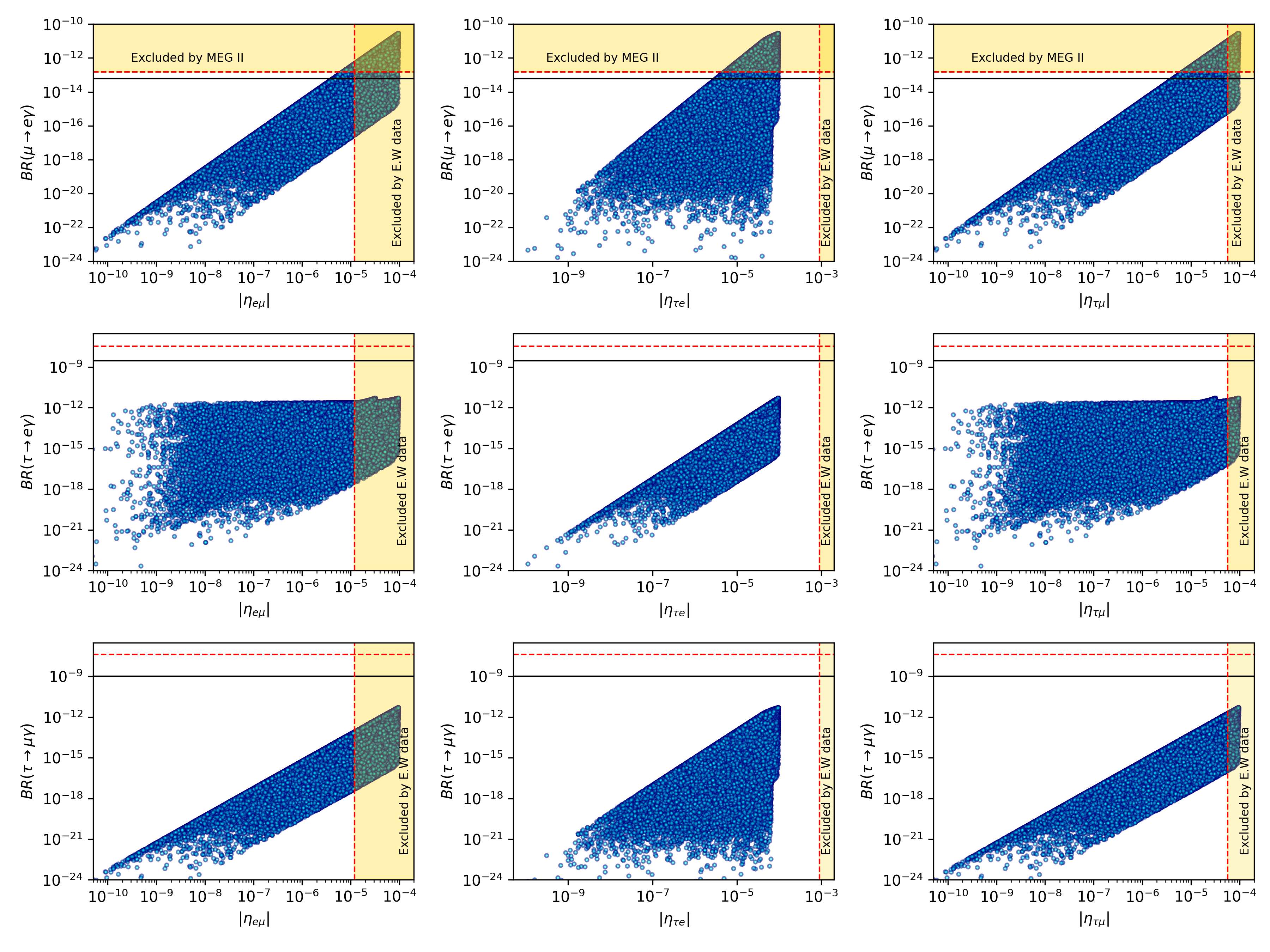}
\caption{\footnotesize \justifying{ Branching ratio for $\ell_{\alpha} \to \ell_{\beta} \gamma$} with $\alpha = \mu, \tau$ and $\beta = e, \mu$ in our model using the cross $\xi$ texture compared with the absolute value of the elements of the mixing matrix $|\eta_{\alpha \beta}|$. The red dotted vertical and horizontal lines represent the current experimental limits, the solid black lines represent future sensitivities while the yellow regions are excluded by the electroweak (E.W) data defined in Eq.~\eqref{eq:29}.}
\label{fig:eta}
\end{figure}

\end{widetext}

Regarding the relationship between non-unitary effects and measurable cLFV decays, Figure~\ref{fig:eta} presents the correlation between $\ell_{\alpha} \to \ell_{\beta} \gamma$ decay rates and the absolute values of the parameters $\eta_{\alpha \beta}$. The $\mu \to e \gamma$ decay shows stronger correlation with $|\eta_{\mu e}|$ than with $|\eta_{\tau e}|$, while a particularly strong correlation appears with $|\eta_{\tau \mu}|$ due to the relations in Eq.~\eqref{eq:28} that originate from the chosen parameterization of the $\xi$ matrix. Similarly, this behavior is also observed in the $\tau \to \mu \gamma$ channel, since the relationships shown in Eq.~\eqref{eq:28} produce a strong correlation with the parameters $|\eta_{\tau \mu}|$ and $|\eta_{\mu e}|$. The $\tau \to e \gamma$ decay exhibits a significant correlation exclusively with $|\eta_{\tau e}|$, as theoretically expected.
\\

In all plots of Figure~\ref{fig:eta}, horizontal dotted red and solid black lines represent current upper bounds and future sensitivity limits for branching ratios, respectively (see Table~\ref{tab:limi}). The yellow horizontal band in the $\mu \to e \gamma$ plot shows parameter space excluded by the MEG II collaboration~\cite{MEGII:2025gzr}, while vertical dotted red lines indicate bounds on non-unitary effects from the $\eta$ matrix shown in Eq.~\eqref{eq:29}. As anticipated, $\mu \to e \gamma$ emerges as the most stringently constrained decay channel due to current experimental limits. The equality of matrix elements $\eta_{\mu e} = \eta_{ \tau \mu}$ leads to an intrinsically correlated behavior between the $\mu \to e \gamma$ and $\tau \to \mu \gamma$ branching ratios, as evident in their corresponding graphical representations.
\\

Through a comprehensive parameter scan combining cLFV branching ratio limits with non-unitarity constraints, we analyze the distribution of generated points shown in Fig.~\ref{fig:eta}. This yields the following upper bounds on the absolute values of the non-unitarity matrix elements:
\begin{equation}\label{cotas}
    \begin{aligned}
        & |\eta_{\mu e}| \lesssim 4.8 \times 10^{-6}, \\
        & |\eta_{\tau e}| \lesssim 2.0 \times 10^{-5}, \\
        & |\eta_{ \tau \mu }| \lesssim 4.8 \times 10^{-6},
    \end{aligned}
\end{equation}
which are required to satisfy the stringent experimental limits from $\mu \to e \gamma$ decays. These results for the non-unitarity matrix elements agree well with those in Ref.~\cite{Blennow:2023mqx}.
\\

Regarding our numerical estimates for the non-unitary parameters $|\eta_{\tau e}|$ and $|\eta_{\tau\mu}|$, we find it is worth emphasizing that there are significant discrepancies between our approximate estimations, given in Eq.~\eqref{eq:20}, and the numerical results shown in Eq.~\eqref{cotas}. Such differences are as large as 2 or 3 orders of magnitude. This arises because the former considered only the experimental bounds on branching ratios, being unconnected to each other, while the later considers, in addition, the constraints from $\mu \to e \gamma$. Both approaches show perfect agreement for $|\eta_{\mu e}|$. 

Using the relations in Eq.~\eqref{eq:28}, we estimate the upper bounds for the entries of the non-unitary matrix $\eta$, expressed in terms of the parameters $\rho_1$ and $\rho_2$ as defined in Eq.~\eqref{eq:27}

\begin{equation}
    |\eta| \lesssim \begin{pmatrix}
        2.0 \times 10^{-5} & 4.8 \times 10^{-6} & 2.0 \times 10^{-5} \\
        4.8 \times 10^{-6} & 1.7 \times 10^{-5} & 4.8 \times 10^{-6} \\
        2.0 \times 10^{-5} & 4.8 \times 10^{-6} & 2.0 \times 10^{-5}
    \end{pmatrix}. 
\end{equation}

%%%%%%%%%%%%%%%%%%%%%%%%%%%%%%

\subsection{Triangle \texorpdfstring{$\xi$}\text{\hspace{2mm}texture}}

The triangle texture is defined by the matrix
\begin{equation}\label{tria}
   \xi = e^{i\phi}\begin{pmatrix}
        \rho_2 & \rho_2 & \rho_1 \\
        \rho_2 & \rho_1 & \rho_1 \\
        \rho_1 & \rho_1 & \rho_1
    \end{pmatrix},
\end{equation}
where $\rho_1$ and $\rho_2$ are real-valued parameters and $\phi$ is a phase. The triangle texture generates a symmetric non-unitary matrix $\eta$, with the following structure:
\begin{equation}
\begin{aligned}
    \eta_{\alpha \beta} &= \left(\frac{\xi \xi^\dagger}{2} \right)_{\alpha \beta} \\
    & = \begin{pmatrix}
        \rho^2_1+2\rho^2_2 & \rho^2_1 + \rho_1 \rho_2 + \rho^2_2 & \rho^2_1 + 2\rho_1 \rho_2 \\
         \rho^2_1 + \rho_1 \rho_2 + \rho^2_2 & 2\rho^2_1+\rho^2_2 & 2\rho^2_1+\rho_1\rho_2 \\
         \rho^2_1 + 2\rho_1 \rho_2 & 2\rho^2_1+\rho_1\rho_2 & 3\rho^2_1
    \end{pmatrix}_{\alpha \beta}.
\end{aligned}
\end{equation}
It is crucial to emphasize that, in contrast to the previous texture, this scenario does not exhibit identical relationships between mixing matrix elements with different flavor indices. Within this parametrization, the $\mu \rightarrow e \gamma$ branching ratio – being a function of $|\eta_{\mu e}|$ – scales proportionally with the following parameter combination:
\begin{equation}
    \mathcal{BR}(\mu \to e \gamma) \propto  (\rho^2_1 + \rho_1 \rho_2 + \rho^2_2 )\times f(m_W, M_N, m_\mu, m_e),
    \label{esoval}
\end{equation}
where $f(m_W, M_N, m_\mu, m_e)$ denotes the same mass-dependent function introduced for the cross-texture scenario. 
\begin{figure}[H]
\centering
\includegraphics[width=0.53\textwidth]{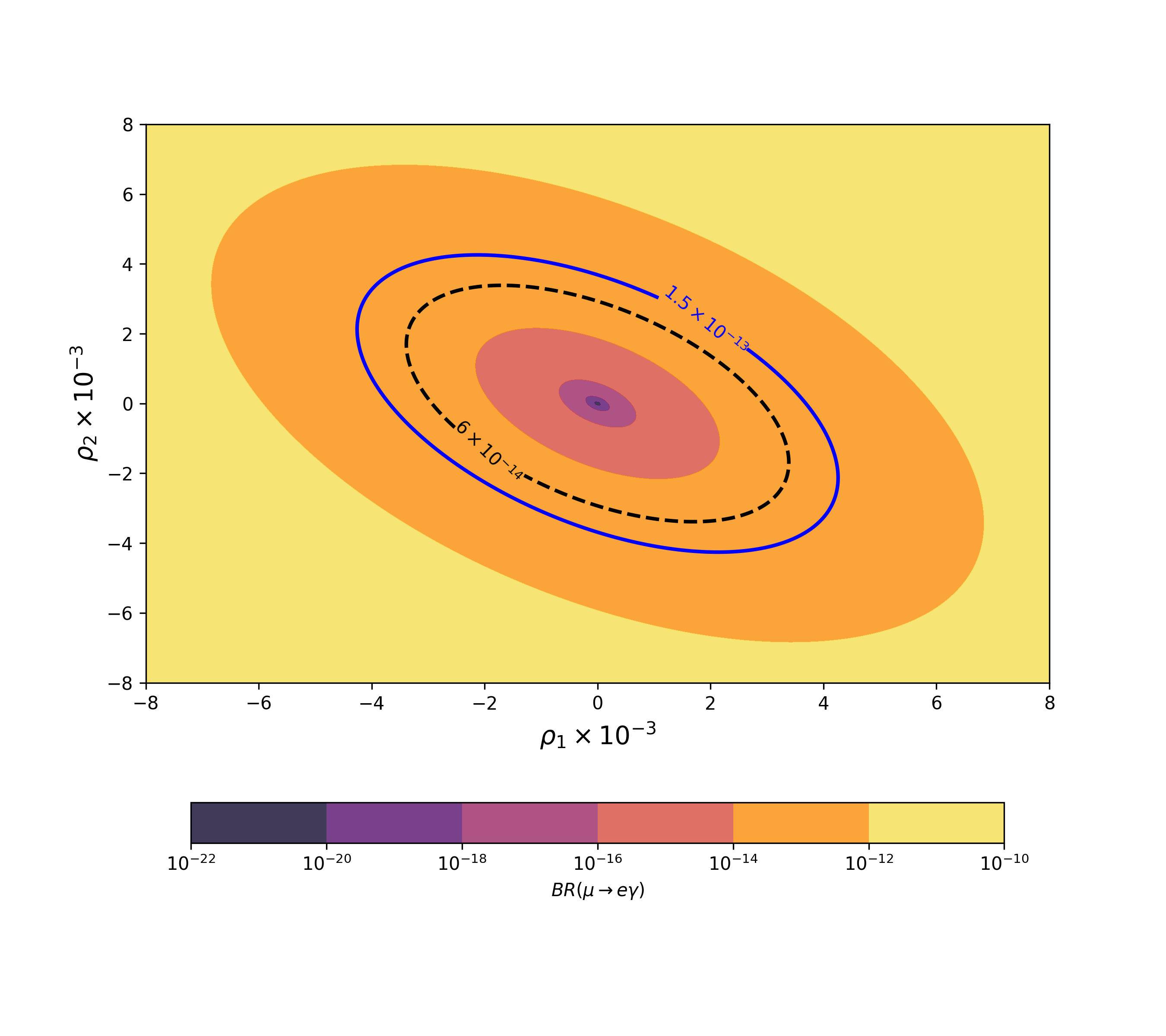}
\caption{\footnotesize \justifying{Branching ratio $\mathcal{BR}(\mu \to e \gamma)$ in the $\rho_1$--$\rho_2$ parameter space. The current experimental upper limit $1.5 \times 10^{-13}$ is represented by the blue solid curve; the projected sensitivity, $6 \times 10^{-14}$, is represented by the black dashed curve (see Table~\ref{tab:limi}).}}
\label{fig:espaciooval}
\end{figure}
Figure~\ref{fig:espaciooval} displays the allowed parameter space in the $(\rho_1, \rho_2)$ plane, calculated under identical constraints to those used in the cross-texture analysis (see Fig.~\ref{fig:espacio} for context). The resulting elliptical distribution of parameter points reflects the quadratic dependence of Eq.~\eqref{esoval} on the combination $\rho_1^2 + \rho_1\rho_2 + \rho_2^2$.
\\

In this texture, the $\ell_{\alpha} \to \ell_{\beta} \gamma$ branching ratios, as functions of the heavy-neutrino mass, exhibit a behavior consistent with the results shown in Figs.~\ref{fig:mue} and \ref{fig:taus}. However, only 45\% of the parameter points in $\mathcal{BR}(\mu \to e\gamma)$ satisfy the current experimental upper limit from MEG II, a direct consequence of modifications in the mixing matrix elements. In contrast, the branching ratios for $\tau \to e\gamma$ and $\tau \to \mu\gamma$ remain largely unaffected by this new parameterization. 
\\

An analysis of the parameter distributions (not shown) yields the following upper bounds on the absolute values of the non-unitarity matrix elements: 
\begin{equation}\label{eq:eta1}
    \begin{aligned}
        & |\eta_{\mu e}| \lesssim 6.8 \times 10^{-6}, \\
        & |\eta_{\tau e}| \lesssim 5.3 \times 10^{-6}, \\
        & |\eta_{\tau \mu }| \lesssim 7.0 \times 10^{-6}.
    \end{aligned}
\end{equation}
The numerical results shown in Eq.~\eqref{eq:eta1} are consistent again with the results of Ref.~\cite{Blennow:2023mqx} and we should point out that they are also consistent with the results obtained using the cross parameterization (see Eq.~\eqref{cotas}), which implies that changing the parameterization does not produce significant changes in the behavior of the $\ell_{\alpha} \to \ell_{\beta} \gamma$ branching ratios or in the elements of the matrix $\eta$. The full $\eta$ matrix as a function of the parameters $\rho_1$ and $\rho_2$ acquires the following numerical values: 
\begin{equation}
    |\eta| \lesssim \begin{pmatrix}
        1.3 \times 10^{-5} & 6.8 \times 10^{-6} & 5.3 \times 10^{-6} \\
        6.8 \times 10^{-6} & 1.3 \times 10^{-5} & 7.0 \times 10^{-6} \\
        7.0 \times 10^{-6} & 5.3 \times 10^{-6} & 1.3 \times 10^{-5}
    \end{pmatrix},
\end{equation}
where the matrix exhibits equal diagonal elements of order $\mathcal{O}(10^{-5})$, while the off-diagonal elements are of order $\mathcal{O}(10^{-6})$. 
\\

\begin{widetext}

\begin{figure}[H]
 \centering
\includegraphics[scale=0.6]{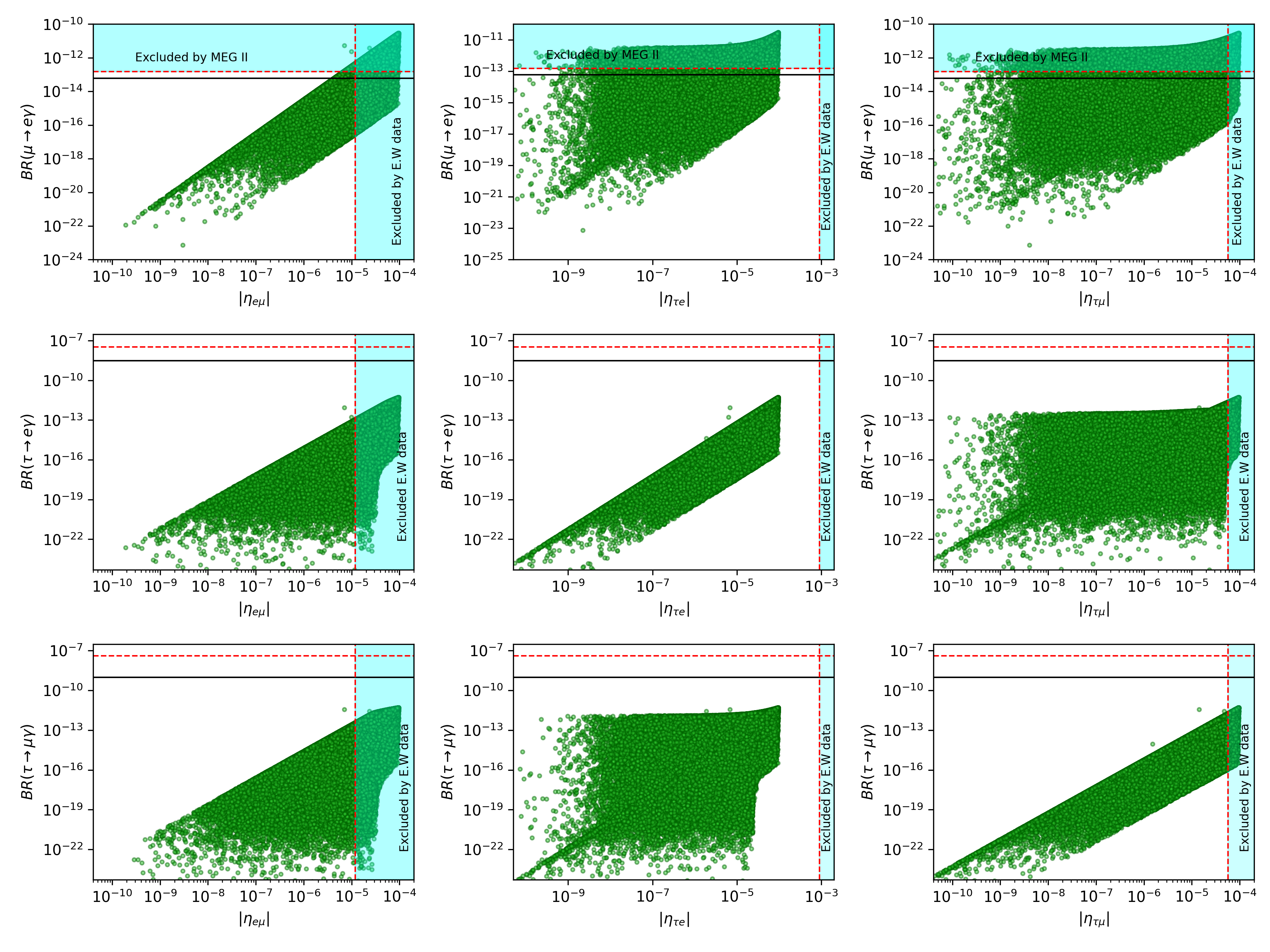}
\caption{\footnotesize \justifying{ Branching ratio for the $\ell_{\alpha} \to \ell_{\beta} \gamma$} with $\alpha = \mu, \tau$ and $\beta = e, \mu$, using the triangle $\xi$ texture compared with the moduli of the elements of the mixing matrix $|\eta_{\alpha \beta}|$. The red dotted vertical and horizontal lines represent the current experimental limits, the solid black lines represent future sensitivities while the blue regions are excluded by the electroweak (E.W) data defined in Eq.~\eqref{eq:29}.}
\label{fig:etaoval}
\end{figure}

\end{widetext}

Figure~\ref{fig:etaoval} depicts the correlation between non-unitary effects and the cLFV branching ratios, analogous to the cross-texture scenario discussed earlier. However, in contrast to the cross-parametrization—where the $\mu \to e \gamma$ and $\tau \to \mu \gamma$ channels depend on distinct elements of the $\eta$ matrix— a straightforward calculation reveals that this scenario has two distinctive features:  
\begin{itemize}
    \item The $\mu \to e \gamma$ decay depends solely on $|\eta_{\mu e}|$, namely,
    \begin{equation}
{\cal BR}(\mu\to e\gamma)\propto\rho_1^2+\rho_1\rho_2+\rho_2^2=|\eta_{\mu e}|.
    \end{equation}
    \item The $\tau \to e \gamma$ and $\tau \to \mu \gamma$ processes correlate exclusively with $|\eta_{\tau e}|$ and $|\eta_{\tau \mu}|$, respectively, exhibiting similar functional behavior, that is,
    \begin{equation}
{\cal BR}(\tau\to e\gamma)\propto\rho_1^2+2\rho_1\rho_2=|\eta_{\tau e}|,
    \end{equation}
    \vspace{0.1cm}
    %%%%%
    \begin{equation}
{\cal BR}(\tau\to \mu\gamma)\propto2\rho_1^2+\rho_1\rho_2=|\eta_{\tau \mu}|.
    \end{equation}
\end{itemize}

\subsection{Diagonal-triangle \texorpdfstring{$\xi$}\text{\hspace{2mm}texture}}

The last texture to be analyzed is the diagonal-triangle texture, distinguished by the $\xi$ matrix form

\begin{equation}\label{tria1}
   \xi = e^{i\phi}\begin{pmatrix}
        \rho_1 & 0 & 0 \\
        \rho_2 & \rho_1 & 0 \\
        \rho_2 & \rho_2 & \rho_1
    \end{pmatrix},
\end{equation}
with $\rho_1$, $\rho_2$ representing real-valued parameters and $\phi$ denoting a phase. This matrix generates a symmetric matrix $\eta$ with the following structure:

\begin{equation}\label{eq:diago}
\begin{aligned}
    \eta_{\alpha \beta} &= \left(\frac{\xi \xi^\dagger}{2} \right)_{\alpha \beta} \\
    & = \begin{pmatrix}
        \rho^2_1 & \rho_1 \rho_2 & \rho_1 \rho_2 \\
        \rho_1 \rho_2 & \rho^2_1+\rho^2_2 & \rho_1\rho_2+\rho^2_2 \\
         \rho_1 \rho_2 & \rho_1\rho_2+\rho^2_2 & \rho^2_1 +2\rho^2_2
    \end{pmatrix}_{\alpha \beta}.
\end{aligned}
\end{equation}
It can be seen that in Eq.~\eqref{eq:diago} the following relationships exist between the elements of the mixing matrix $\eta$:

\begin{equation}
    \eta_{e \mu} = \eta_{ \mu e} = \eta_{e \tau} = \eta_{\tau e }.
\end{equation}
As a particular case, this parameterization gives rise to symmetry equations that relate the elements $\eta_{\mu e} = \eta_{\tau e}$. This result is similar to the equalities of the cross parameterization given in Eq.~\eqref{eq:28} that related the elements $\eta_{\mu e} = \eta_{\tau \mu}$. Within this framework, the branching ratio for the $\mu \rightarrow e \gamma$ decay exhibits proportionality to the particular parameter combination

\begin{equation}
    \mathcal{BR}(\mu \to e \gamma) \propto  \rho_1 \rho_2\times f(m_W, M_N, m_\mu, m_e),
    \label{escruz}
\end{equation}
where $f(m_W, M_N, m_\mu, m_e)$ denotes the same mass-dependent function introduced for the cross-texture and triangle-texture scenarios.

\begin{figure}[H]
\centering
\includegraphics[width=0.53\textwidth]{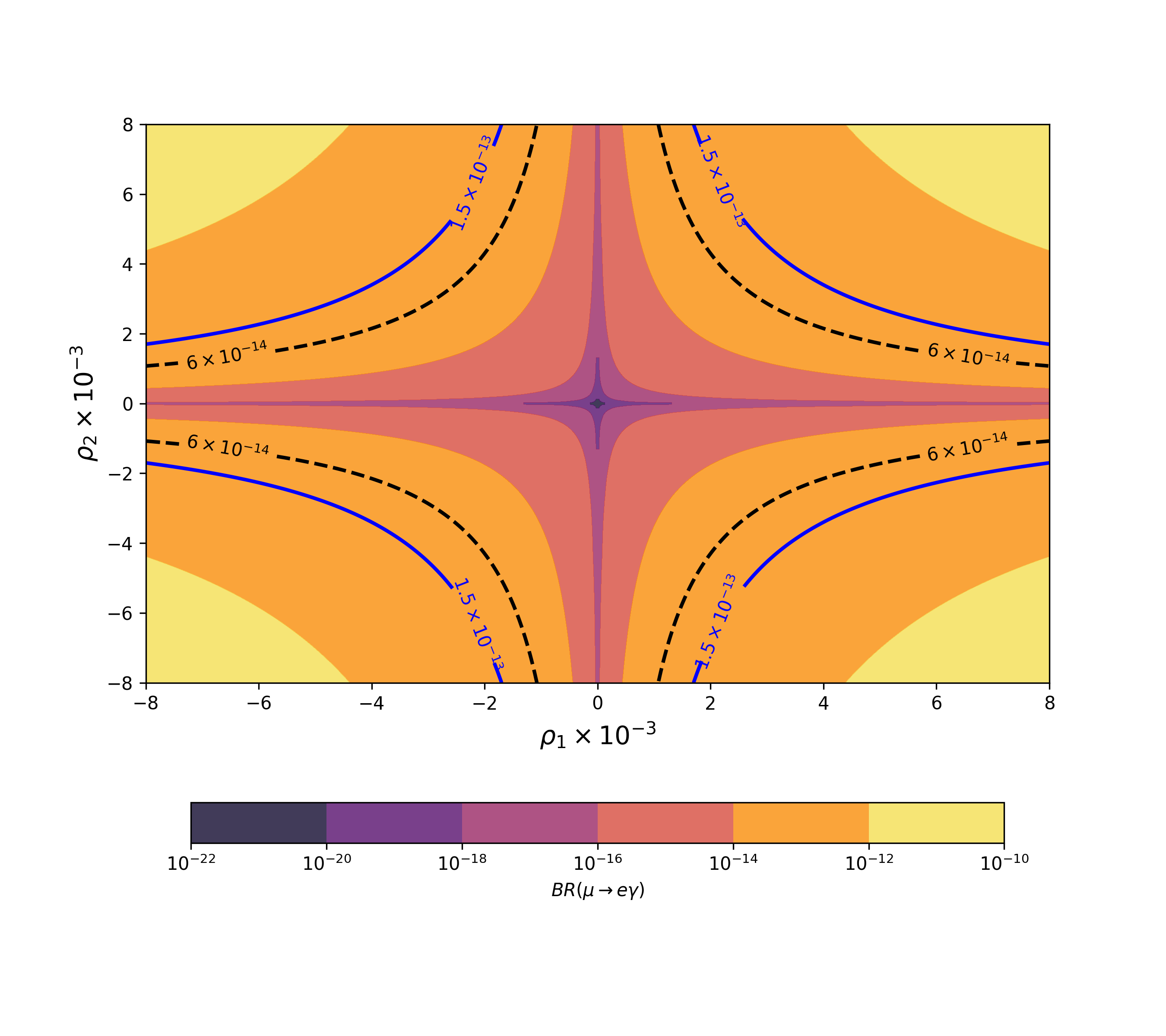}
\caption{\footnotesize \justifying{Branching ratio $\mathcal{BR}(\mu \to e \gamma)$ in the $\rho_1$--$\rho_2$ parameter space. The current experimental upper limit $1.5 \times 10^{-13}$ is represented by the blue solid curve; the projected sensitivity, $6 \times 10^{-14}$, is represented by the black dashed curve (see Table~\ref{tab:limi}).}}
\label{fig:espaciocruz}
\end{figure}

Figure~\ref{fig:espaciocruz} shows the allowed parameter space in the $(\rho_1, \rho_2)$ plane, computed by using the same constraints as in the cross-texture analysis (see Fig.~\ref{fig:espacio}). The resulting hyperbolic distribution of parameter points emerges from the dependence of Eq.~\eqref{escruz} on the product $\rho_1\rho_2$. These hyperbolas share common asymptotes in $\rho_1 = 0$ and $\rho_2 = 0$, with their characteristic shape revealing an inverse relationship between $\rho_1$ and $\rho_2$ in determining the branching ratio. This parameter space is very similar to that displayed in Fig.~\ref{fig:espacio}, showing that different parameterizations of the matrix $\xi$ can lead to similar parameter spaces and results.\\

Within this parametrization, the $\ell_{\alpha} \to \ell_{\beta} \gamma$ branching ratios as functions of heavy-neutrino mass exhibit a behavior identical to that presented in Figs.~\ref{fig:mue} and~\ref{fig:taus}, maintaining the same order of magnitude as previous results. The key difference lies in the $\mathcal{BR}(\mu \to e\gamma)$ distribution, where 71\% of parameter points now satisfy the current experimental upper limit from MEG II. Notably, the $\tau \to e\gamma$ and $\tau \to \mu\gamma$ branching ratios remain essentially unchanged. Our analysis (not shown) with this parameterization yields the following upper bounds on the non-unitarity matrix elements:

\begin{equation}\label{eq:eta2}
    \begin{aligned}
        & |\eta_{\mu e}| \lesssim 3.7 \times 10^{-6}, \\
        & |\eta_{\tau e}| \lesssim 3.7 \times 10^{-6}, \\
        & |\eta_{\tau \mu}| \lesssim 7.8 \times 10^{-6}.
    \end{aligned}
\end{equation}

Our numerical results in Eq.~\eqref{eq:eta2} agree with both the findings of Ref.~\cite{Blennow:2023mqx} and previous parameterizations (cf. Eqs.~\eqref{cotas} and~\eqref{eq:eta1}). This consistency demonstrates that neither the mixing matrix elements $\eta$ nor the $\ell_{\alpha} \to \ell_{\beta} \gamma$ branching ratios undergo significant modifications under different $\xi$ matrix parameterizations. \\

This universal behavior stems from the common analytical structure shared by all decay channels. The primary distinction between channels arises from the parameters $|\eta_{\alpha \beta}|$, which are constrained by the stringent experimental bounds on $\mathcal{BR}(\mu \to e\gamma)$. The elements of the $\eta$ matrix as functions of $\rho_1$ and $\rho_2$ are given by

\begin{equation}
    |\eta| \lesssim \begin{pmatrix}
        7.0 \times 10^{-6} & 3.7 \times 10^{-6} & 3.7 \times 10^{-6} \\
        3.7 \times 10^{-6} & 1.4 \times 10^{-5} & 7.9 \times 10^{-6} \\
        3.7 \times 10^{-6} & 7.9 \times 10^{-6} & 2.1 \times 10^{-5}
    \end{pmatrix}.
\end{equation}

\begin{widetext}

\begin{figure}[H]
 \centering
\includegraphics[scale=0.6]{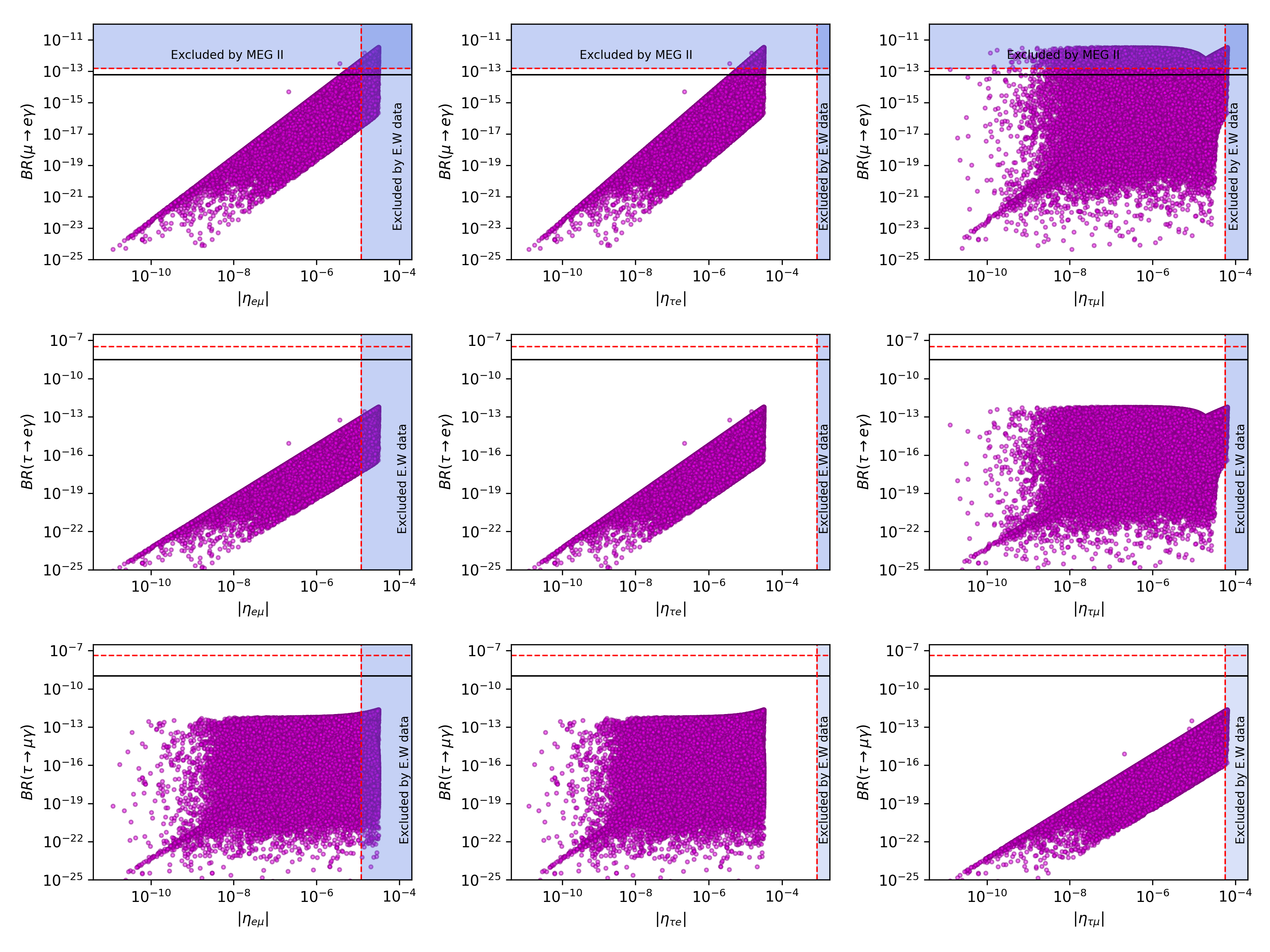}
\caption{\footnotesize \justifying{ Branching ratio for the $\ell_{\alpha} \to \ell_{\beta} \gamma$} with $\alpha = \mu, \tau$ and $\beta = e, \mu$, using the diagonal-triangle $\xi$ texture compared with the absolute value of the elements of the mixing matrix $|\eta_{\alpha \beta}|$. The red dotted vertical and horizontal lines represent the current experimental limits, the solid black lines represent future sensitivities while the blue regions are excluded by the electroweak (E.W) data defined in Eq.~\eqref{eq:29}.}
\label{fig:etacruz}
\end{figure}

\end{widetext}

Figure~\ref{fig:etacruz} displays the connection between non-unitary effects and observable cLFV decays, analogously to the different textures discussed previously. In this parameterization, we see that the decays $\mu \to e\gamma$ and $\tau \to e\gamma$ show a strong correlation with the two parameters $|\eta_{\mu e}|$ and $|\eta_{\tau e}|$, while the channel $\tau \to \mu \gamma$ is only correlated with the parameter $|\eta_{\tau \mu}|$. This behavior comes from the relationships shown in Eq.~\eqref{eq:eta2} and is very similar to the results of Figure~\ref{fig:eta} in the cross $\xi$ parameterization.

\section{Summary and conclusion}\label{con}

The fundamental nature of neutrinos - whether Dirac or Majorana particles - remains one of the most pressing questions in particle physics. While many beyond-SM scenarios predict measurable cLFV processes, these are strongly suppressed in the minimal extension of the SM characterized by massive Dirac neutrinos, due to the tiny neutrino masses. \\

In this work, we have investigated a radiative seesaw variant where light-neutrino masses are naturally suppressed, requiring quasi-degenerate heavy neutrinos with masses in the TeV range. We derived complete loop-level analytical expressions for the radiative decays $\ell_\alpha \to \ell_\beta\gamma$ and performed a detailed phenomenological analysis for different textures of the heavy–light mixing matrix. Our results demonstrate that, in this framework, the decay $\mu \to e\gamma$ is the most constraining cLFV channel at present, owing to its stringent experimental bound $\mathrm{BR}(\mu \to e\gamma) \lesssim 1.5 \times10^{-13}$.
\\

We have derived complete loop-level analytical expressions for $\ell_{\alpha} \to \ell_{\beta} \gamma$ decays, finding the characteristic structure:
\begin{equation}
    \mathcal{BR}\left( \ell_{\alpha} \to \ell_{\beta} \gamma  \right) \approx C_{\alpha \beta} \Big| \eta_{\alpha \beta} \left( H_{N}\left( y_i \right)-H_{\nu}\left( x_i \right) \right) \Big|^2,
\end{equation}
where $H_{N}$ and $H_{\nu}$ represent the ultraviolet-finite contributions from heavy and light neutrinos respectively, expressed in terms of electric and magnetic form factors. The amplitude $\Gamma\left(\ell_\alpha \rightarrow \ell_\beta \gamma\right)$ satisfies the Ward identity and reproduces SM results if heavy neutrinos are assumed to be absent. In the limit as $\frac{M_N^2}{m_W^2}\to\infty$ and $\frac{m_{\nu}^2}{m_W^2}\to 0$, the GIM suppression is lifted, leading to the simplified form in the large mass limit:
\begin{equation}
    \mathcal{BR}\left( \ell_{\alpha} \to \ell_{\beta} \gamma  \right) \approx \frac{3 \alpha_{\rm w}}{2\pi}|\eta_{\alpha \beta}|^2.
\end{equation}\\

However, let us emphasize that for our numerical evaluation such an approximation has not been used, but the complete expression for the branching ratio has been taken instead, in which case the range of masses used for heavy neutrinos does not cause the contribution of $H_{N}\left( y_i \right)-H_{\nu}\left( x_i \right)$ to have constant values, allowing variations in the branching ratios depending on the mass of the heavy neutrino. For each parameterization of the matrix $\xi$ the following values were found for the off-diagonal elements of the non-unitary matrix $\eta$.\\

\begin{table}[h]
\centering
\footnotesize
\setlength{\tabcolsep}{4pt}
\begin{tabular}{|c|c|c|c|}
\hline
\textbf{Parameter} & \textbf{Cross} & \textbf{Triangle} & \textbf{Diag-Tri} \\ 
\hline
$\eta_{ee}$ & $2.0\times 10^{-5}$ & $1.3\times 10^{-5}$ & $7.0\times 10^{-6}$\\
\hline
$\eta_{\mu\mu}$ & $1.7\times 10^{-5}$ & $1.3\times 10^{-5}$ & $1.4\times 10^{-5}$ \\
\hline
$\eta_{\tau\tau}$ & $2.0\times 10^{-5}$ & $1.3\times 10^{-5}$ & $2.1\times 10^{-5}$ \\
\hline
$|\eta_{\mu e}|$ & $4.8\times 10^{-6}$ & $6.8\times 10^{-5}$ & $3.7\times 10^{-6}$ \\
\hline
$|\eta_{\tau e}|$ & $2.0\times 10^{-5}$ & $5.3\times 10^{-6}$ & $3.7\times 10^{-6}$ \\
\hline
$|\eta_{\tau \mu}|$ & $4.8\times 10^{-6}$ & $7.0\times 10^{-6}$ & $7.9\times 10^{-6}$ \\
\hline
\end{tabular}
\caption{\justifying{Bounds on non-unitarity parameters $\eta_{\alpha\beta}$ for different $\xi$ matrix textures.}}
\label{tab:textable}
\end{table}

Initially, we compare each branching ratio $\mathcal{BR}(\ell_\alpha \to \ell_\beta \gamma)$, separately and independently, with its respective experimental bounds from Table~\ref{tab:limi}, deriving the limits $|\eta_{\mu e}| \lesssim 10^{-6}$, $|\eta_{\tau e}| \lesssim 10^{-3}$, and $|\eta_{\tau \mu}| \lesssim 10^{-3}$ under the non-general assumption that heavy neutrino masses are very large, namely, that $M_{N_i} \gg m_W$ holds. These bounds are obtained without assuming any relationship between the different flavor channels. When performing the numerical estimation considering the physical range of heavy neutrino mass $M_{N_i} \in [10, 1500]$ GeV, we impose the most stringent constraint from $\mu \to e\gamma$ to ensure all parameter combinations $(\rho_1, \rho_2)$ simultaneously satisfy $\mathcal{BR}(\mu \to e\gamma) < 1.5 \times 10^{-13}$, $\mathcal{BR}(\tau \to e\gamma) < 3.3 \times 10^{-8}$, and $\mathcal{BR}(\tau \to \mu\gamma) < 4.2 \times 10^{-8}$. This procedure yields the unified constraints shown in Table~\ref{tab:textable}, with all $|\eta_{\alpha\beta}|$ parameters constrained to orders of magnitude between $10^{-5}$ and $10^{-6}$ for the range of mass considered.\\

Our comprehensive analysis considered four different $\xi$ matrix parameterizations: quasi-diagonal texture, cross-texture, triangle texture, and diagonal-triangle texture. While parameterizations proportional to identity or satisfying $\xi \xi^{\dagger} \sim \textbf{1}_3$ yield highly suppressed branching ratios, the other three parameterizations produced results similar to each other across the explored parameter space ($M_N \in [10,1500]$ GeV, $\rho_{1,2} \in [-8,8]\times 10^{-3}$). In all cases, the $\mu \to e\gamma$ channel emerges as the most sensitive probe. This behavior is not accidental: the branching ratios are controlled by $|\sum^{3}_{i=1}\xi_{\alpha i} \xi^{*}_{\beta i} G(x_i,y_i)|^2$, where the matrix elements $\eta_{\alpha \beta}=\big( \frac{1}{2}\xi\xi^\dag \big)_{\alpha\beta}$ play a crucial role during the numerical evaluation, while the parameterizations impose relations among the $\eta$ entries, which propagate the stringent $\mu\to e\gamma$ constraint into the tau sector. As a consequence, only $\mu \to e\gamma$ can reach the present and near-future
experimental sensitivity, while $\tau \to \ell \gamma$ processes remain suppressed by construction of the model.\\ 

The analysis of the $\rho_1, \rho_2$ parameter space revealed distinctive allowed regions for each $\xi$ texture: hyperbolas suggesting an inverse relationship between parameters for the cross-texture and for the diagonal-triangle texture, whereas the triangle texture produced an elliptical boundary constraint. Moreover, the $\eta$ textures resulting from the cross and from the diagonal-triangle $\xi$ textures involve relations among the non-unitarity-matrix entries, a feature which, by contrast, is not found in the $\eta$ matrix corresponding to the triangle $\xi$ texture. Let us emphasize that in the three cases, corresponding to these three $\xi$ textures, the stringent constraints on the non-unitarity $|\eta_{\mu e}|$ parameter induce a noticeable suppression on the tau-decay processes.
\\

The $\mathcal{BR}(\ell_{\alpha} \rightarrow \ell_\beta \gamma )$ versus $M_N$ plots exhibit asymptotic behavior that allows the muon process to reach current and future experimental sensitivity, in contrast to the tau channels which remain three orders of magnitude below these bounds - a pattern consistent across all three studied textures. The $\mathcal{BR}(\ell_{\alpha} \rightarrow \ell_\beta \gamma )$ versus $\eta_{\alpha \beta}$ plots reveal different correlations between these elements according to the symmetry conditions of the $\eta$ matrix entries.\\

Comparison with current experimental limits shows that $\mu \to e \gamma$ emerges as the most promising discovery channel, while $\tau$ decays ($\sim 10^{-13}$) remain beyond current (Belle and BaBar) and projected sensitivities (Belle II). The non-unitarity constraints require $|\eta_{\mu e}| \lesssim 10^{-6}$, with corresponding suppression for $|\eta_{\tau e}|$ and $|\eta_{\tau \mu}|$. These results demonstrate that our radiative seesaw framework naturally explains neutrino mass suppression while remaining consistent with current cLFV bounds and offering testable predictions for future experiments.\\

Although other channels---such as $\mu \to 3e$ and $\mu{-}e$ conversion---can be correlated with $\mu \to e\gamma$ in this and many other NP scenarios, their full treatment requires inclusion of additional loop topologies (penguin and box diagrams) and, in the case of $\mu{-}e$ conversion, nuclear form factors. These calculations, while standard in other contexts, require nontrivial adaptations to our radiative seesaw variant and are beyond the scope of the present paper. We are currently carrying out this extended analysis, which will be reported in a dedicated follow-up work aimed at comparing all three processes within the same model.
\\

Our present findings therefore serve as a first, focused step in exploring the loop-level structure of radiative cLFV in this model, providing benchmark constraints and highlighting the parameter regions where future experiments---such as MEG~II---may be sensitive to new physics.\\

%Nowadays, radiative decays $\ell_{\alpha} \rightarrow \ell_\beta \gamma$ have dominated the experimental bounds of cLFV processes. However, the whole set of possible cLFV decays also includes three-body decays $\ell_{\alpha} \to 3 \ell_{\beta}$, $\mu - e$ nuclei conversion, Higgs and $Z$ boson decays. So far, the best experimental limits come from the process $\mu \to e \gamma$, though future experimental limits for the channels $\mu \to ee\bar{e}$ and $\mu-e$ nuclei conversion with aluminum (titanium) will be of order of $10^{-16}$ and $3 \times 10^{-17}$ ($10^{-18}$)~\cite{COMET:2025sdw,Alekou:2013eta}, respectively, in which case these two processes are expected to become the main references in the search for NP linked to cLFV. Such quite relevant processes are being addressed by the authors of the present work in an ongoing investigation.

\section*{Acknowledgements}
\noindent
The authors acknowledge financial support from SECIHTI (México). The work of E. R. and M. S. is funded by \textit{Estancias Posdoctorales por México}, H. V. acknowledges financial support from SECIHTI ``\textit{Becas Nacionales para estudios de Posgrado 2024-2}" program, with support number 4037802.

\appendix
\section{Loop functions}\label{Ape}
The Passarino–Veltman tensor reduction~\cite{Passarino,DEVARAJ} method for evaluating loop integrals is widely used in particle physics. A general triangle Feynman diagram is as follows:

\begin{figure}[H]
\centering
\includegraphics[width=0.3\textwidth]{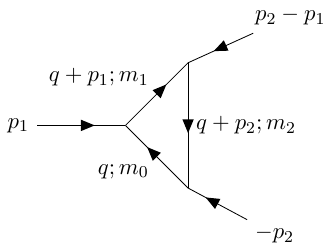}
\caption{\footnotesize \centering{General penguin diagram.}}
%\label{}
\end{figure}

We can express the general form of a three-point tensor integral as follows:
 
\begin{widetext}
\begin{equation}
    \frac{i}{16 \pi^2} \left[C_0, C_{\mu}, C_{\mu \nu} \right] (arg) = \mu^{4-D}\int \frac{d^Dq}{(2\pi)^D} \frac{\left[1,q_{\mu},q_{\mu}q_{\nu} \right]}{(q^2-m^2_0)\left[ (q+p_1)^2-m^2_1  \right]\left[ (q+p_2)^2-m^2_2  \right]},
\end{equation}
\end{widetext}
here, the external momenta are $ p_1$ and $p_2$. The arguments of the Passarino-Veltman functions $\mathcal{C}_{0}$, $\mathcal{C}_{\mu}$, and $\mathcal{C}_{\mu\nu}$ depend on the invariant quantities $ (arg)=(p_1^2, Q^2, p_2^2; m_0^2, m_1^2, m_2^2)$, where $Q = p_1 - p_2$. The specific arguments relevant to this work are as follows:

\begin{equation}
    C\left( m^2_{\alpha},0,m^2_{\beta}; m_{n_i}, m_W,m_W \right) \equiv C,
\end{equation}
where $q^2 = 0$. After computing the Feynman diagrams shown in Fig.~\ref{fig:diagramas} with the \textsc{Package-X}~\cite{Patel_2015,Patel:2016fam}, we find that the analytic expressions for the MDFF and the EDFF form factors are: 

\begin{widetext}
\begin{equation}\label{eq:ap1}
	\begin{split}
		F_1 &=-\frac{2 \pi \alpha_{\rm w}}{s^2_{\rm w} } \sum^6_{i=1}{\cal B}_{\alpha n_i}{\cal B}^{*}_{\beta n_i} \left(\frac{i}{2 m^2_W} \right) \Big[ m_{\alpha} \left(m_{\alpha}m_{\beta}+m^2_{n_i} + (D-2)m^2_W \right) C_{11}+ m_{\beta} \left(m_{\alpha}m_{\beta}+m^2_{n_i} + (D-2)m^2_W \right) C_{22} \\
		& + \left( m^2_{\alpha}m_{\beta}+ m^2_{n_i}(m_{\beta}+2m_{\alpha})+m^2_W\left(m_{\alpha}(D-4)-2m_{\beta} \right)  \right)C_1 + m^2_{n_i}(m_{\alpha}+m_{\beta})C_0 \\
		& + \left( m_{\alpha}m^2_{\beta}+ m^2_{n_i}(m_{\alpha}+2m_{\beta})+m^2_W\left(m_{\beta}(D-4)-2m_{\alpha} \right)  \right)C_2 + (m_{\alpha}+m_{\beta}) \left(m_{\alpha}m_{\beta}+m^2_{n_i} + (d-2)m^2_W \right)C_{12} \Big], \\
		F_2 &= -\frac{2 \pi \alpha_{\rm w}}{s^2_{\rm w} } \sum^6_{i=1}{\cal B}_{\alpha n_i}{\cal B}^{*}_{\beta n_i} \left(\frac{i}{2 m^2_W} \right) \Big[  m_{\alpha} \left(-m_{\alpha}m_{\beta}+m^2_{n_i} + (D-2)m^2_W \right) C_{11}-m_{\beta} \left(-m_{\alpha}m_{\beta}+m^2_{n_i} + (D-2)m^2_W \right)C_{22}  \\
		& + \left( -m^2_{\alpha}m_{\beta}+ m^2_{n_i}(2m_{\alpha}-m_{\beta})+m^2_W\left(m_{\alpha}(D-4)+2m_{\beta} \right)  \right)C_1 + m^2_{n_i}(m_{\alpha}-m_{\beta})C_0 \\   
		& + \left( m_{\alpha}m^2_{\beta}+ m^2_{n_i}(m_{\alpha}-2m_{\beta})-m^2_W\left(m_{\beta}(D-4)+2m_{\alpha} \right)  \right)C_2 + (m_{\alpha}-m_{\beta}) \left(-m_{\alpha}m_{\beta}+m^2_{n_i} + (D-2)m^2_W \right)C_{12}      \Big].
	\end{split}
\end{equation}
\end{widetext}

Since the form factors in Eq.~\eqref{eq:ap1} are exclusively expressed in terms of 3-point UV-finite functions, we can safely pick the limit $D \to 4$ and use the \textsc{Package-X} to do an analytical evaluation. We expand around the lepton masses throughout the evaluation while preserving the first-order contributions, which yields the following contribution of the PaVe functions:

 \begin{equation}\label{eq:ap2}
        \begin{aligned}
            &C_0 = \frac{1}{m^2_W}  \left[ \frac{-1}{(1-x_i)} - \frac{x_i\ln(x_i)}{(1-x_i)^2} \right] + \mathcal{O}(m^2_{\alpha},m^2_{\beta}), \\
            & C_1 = C_2 = \frac{1}{m^2_W} \left[ \frac{1-3x_i}{4(1-x_i)^2} - \frac{x^2_i \ln(x_i)}{2(1-x_i)^3}  \right] + \mathcal{O}(m^2_{\alpha},m^2_{\beta}), \\ 
            & C_{11} = C_{22} = 2C_{12} = \frac{1}{m^2_W} \left[ \frac{7x_i-11x^2_i-2}{18(1-x_i)^3}-\frac{x^3_i\ln(x_i)}{3(1-x_i)^4}  \right]+ \mathcal{O}(m^2_{\alpha},m^2_{\beta}),
        \end{aligned}
    \end{equation}
Using the relations between PaVe functions given in Eq.~\eqref{eq:ap2} and keeping the expressions to the order $\mathcal{O}(m_{\alpha},m_{\beta})$, the Eq.~\eqref{eq:ap1} can be simplified as:
\begin{widetext}
 \begin{equation}
        \begin{aligned}
            & F_1 = \frac{ \alpha_{\textrm{w}}(m_{\alpha}+ m_{\beta})}{32 \pi s^2_{\textrm{w}}m^2_W} \sum^6_{i=1} \mathcal{B}_{\alpha n_i}\mathcal{B}^*_{\beta n_i}\left[3 (m^2_{n_i} + 2m^2_W)C_{11} + 2(3m^2_{n_i} -2m^2_W)C_1 + 2m^2_{n_i} C_0  \right], \\
            & F_2 = \frac{\alpha_{\textrm{w}}(m_{\alpha}- m_{\beta})}{32 \pi s^2_{\textrm{w}}m^2_W} \sum^6_{i=1} \mathcal{B}_{\alpha n_i}\mathcal{B}^*_{\beta n_i}\left[3 (m^2_{n_i} + 2m^2_W)C_{11} + 2(3m^2_{n_i} -2m^2_W)C_1 + 2m^2_{n_i} C_0  \right].
        \end{aligned}
    \end{equation}
\end{widetext}
Finally, the form factors can be written using Eq.~\eqref{eq:ap2} resulting in:

\begin{equation}
\begin{aligned}
    F_1 &=  - \frac{ \alpha_{\rm w} \left( m_{\alpha}+m_{\beta} \right)}{16\pi s^2_{\rm w} m^2_W} \sum^{6}_{i=1} {\cal B}_{\alpha n_i }{\cal B}^{*}_{\beta n_i } \Big[ \frac{5}{6}- \frac{3x_i-15x^2_i-6x^3_i}{12(1-x_i)^3} \\
    & + \frac{3x^3_i}{2(1-x_i)^4}\ln(x_i) \Big],
\end{aligned}
\end{equation}

\begin{equation}
\begin{aligned}
    F_2 &=  - \frac{ \alpha_{\rm w} \left( m_{\alpha}-m_{\beta} \right)}{16\pi s^2_{\rm w} m^2_W} \sum^{6}_{i=1} {\cal B}_{\alpha n_i }{\cal B}^{*}_{\beta n_i } \Big[ \frac{5}{6}- \frac{3x_i-15x^2_i-6x^3_i}{12(1-x_i)^3} \\
    & + \frac{3x^3_i}{2(1-x_i)^4}\ln(x_i) \Big],
\end{aligned}
\end{equation}

with

\begin{equation}
    x_i = \frac{m^2_{n_{i}}}{m^2_W}, \hspace{3mm} m_{n_i} = \{ m_{\nu_{i}},M_{N_{i}} \}.
\end{equation}
Although it may seem that the contribution of the form factors present a discontinuity at point $x_i=1$, it can be proven that it is a removable singularity and that the function $H(x_i)$ defined in Eq.\eqref{eq:10} is well behaved in the mass interval chosen for the numerical analysis. The case $x_i<1$ occurs when $m_{n_i}<m_W$ and under this condition the dominant term of the function $H(x_i)$ is the constant term $5/6$. This means that in the limit of very small neutrino masses, a finite value is obtained:
\begin{equation}
    \lim_{x \to 0} H(x_i) = \frac{5}{6}.
\end{equation}
The case $x_i>1$ occurs when $m_{n_i}>m_W$ and under this condition the logarithmic term of the function $H(x_i)$ becomes zero and the rational term tends to $-1/2$. This means that in the limit of very large neutrino masses, a finite value is obtained:
\begin{equation}
\lim_{x \to \infty} H(x_i) = \frac{1}{3}.
 \end{equation}
The case of $x_i = 1$ it occurs when $m_{n_i} =m_W$ and it seems that there is a singular point due to the denominators of the function $H(x_i)$, however, this singularity is removable, since the rational and the logarithmic terms contain the same poles up to order 3 which causes the singular terms to cancel each other out giving a finite value at this point. To visualize this, we can make the following change of variable $h = 1-x_i$, then $x \to 1-h$. Now we use the Maclaurin series expansion of the logarithm:
\begin{equation}
    \ln(1-h) = -h - \frac{h^2}{2}-\frac{h^3}{3}-\frac{h^4}{4} + \mathcal{O}(h^5)
 \end{equation}
The limit of the function $H(1-h)$ can be taken when $h\to 0$:
\begin{widetext}
    \begin{equation}
    \begin{aligned}
         & \lim_{h \to 0} H(1-h) = \lim_{h \to 0} \left( \frac{5}{6} - \frac{3(1-h)-15(1-h)^2-6(1-h)^3}{12h^3} + \frac{3(1-h)^3 \ln(1-h)}{2h^4} \right) \\
         & = \lim_{h \to 0} \left( \frac{5}{6} + \frac{3}{2h^3} - \frac{15}{4h^2} + \frac{11}{4h}-\frac{1}{2} + \frac{3}{8} - \frac{3}{2h^3} + \frac{15}{4h^2} - \frac{11}{4h} + \frac{3h}{8} - \frac{5h^2}{8}+\frac{3h^3}{8}  \right) \\
         & = \frac{17}{24},
    \end{aligned}
\end{equation}
\end{widetext}
therefore, the value of the function $H(x_i)$ is well defined for all intervals $x_i<1$, $x_i = 1$ and $x_i>1$.

\bibliographystyle{apsrev4-1}
\bibliography{LFV}

\end{document}